\documentclass[aps,prl,twocolumn,showpacs,superscriptaddress, preprintnumbers, hyperref]{revtex4-1}
\usepackage{latexsym}
\usepackage{amssymb}
\usepackage{graphicx}
\usepackage{amsmath}
\usepackage{bm}
\usepackage[colorlinks,
          linkcolor=black,
            citecolor=black,
            urlcolor=blue
           ]{hyperref}
\usepackage{verbatim}
\usepackage{tabularx}
\usepackage{mathrsfs}
\usepackage{extarrows}
\usepackage{comment}
\usepackage{mathtools,slashed}
\usepackage{soul}
\usepackage[toc,page]{appendix}
\usepackage[vcentermath]{youngtab}
\usepackage{multirow}
\usepackage{atbegshi,picture}
\usepackage{lipsum}
\usepackage{bbm}

\usepackage{ulem}
\usepackage{pifont}

\hypersetup{colorlinks=true,linkcolor=blue,citecolor=blue,urlcolor=blue}

\makeatletter
\newcommand{\SMtableofcontents}{\@starttoc{sm}} 
\newcommand{\SMredirecttoc}{%
  \let\SM@oldaddcontentsline\addcontentsline
  \renewcommand{\addcontentsline}[3]{\SM@oldaddcontentsline{sm}{##2}{##3}}%
}
\newcommand{\SMrestoretoc}{\let\addcontentsline\SM@oldaddcontentsline}
\makeatother

\begin{document}

\title{Non-Hermitian free-fermion critical systems and logarithmic conformal field theory}

\author{Iao-Fai Io}
\email{d11222002@ntu.edu.tw\\}
\affiliation{Department of Physics and Center for Theoretical Physics, National Taiwan University, Taipei 10607, Taiwan}	
\affiliation{RIKEN Center for Interdisciplinary Theoretical and Mathematical Sciences (iTHEMS), Wako 351-0198, Japan}

\author{Fu-Hsiang Huang} 
\affiliation{Department of Physics and Center for Theoretical Physics, National Taiwan University, Taipei 10607, Taiwan}

\author{Chang-Tse Hsieh}         
\email{cthsieh@phys.ntu.edu.tw\\}
\affiliation{Department of Physics and Center for Theoretical Physics, National Taiwan University,
Taipei 10607, Taiwan}
\affiliation{Physics Division, National Center for Theoretical Science, National Taiwan University,
Taipei 10607, Taiwan}
\affiliation{Center for Quantum Science and Engineering, National Taiwan University, Taipei 10617, Taiwan}

\date{\today}

\begin{abstract}
Conformal invariance often accompanies criticality in Hermitian systems. However, its fate in non-Hermitian settings is less clear, especially near exceptional points where the Hamiltonian becomes non-diagonalizable. Here we investigate whether a 1+1-dimensional gapless non-Hermitian system can admit a conformal description, focusing on a PT-symmetric free-fermion field theory. Working in the biorthogonal formalism, we identify the conformal structure of this theory by constructing a traceless energy-momentum tensor whose Fourier modes generate a Virasoro algebra with central charge $c=-2$. This yields a non-Hermitian, biorthogonal realization of a logarithmic conformal field theory, in which correlation functions exhibit logarithmic scaling and the spectrum forms Virasoro staggered modules that are characterized by universal indecomposability parameters. We further present a microscopic construction and show how the same conformal data (with finite-size corrections) can be extracted from the lattice model at exceptional-point criticality, thereby supporting the field-theory prediction.
\end{abstract}

\maketitle

\textit{Introduction---}
Non-Hermitian systems arise naturally in a broad range of physical contexts, including open quantum systems, driven-dissipative systems, and wave dynamics with gain and loss, where the exchange of energy or information with the environment is essential. Such systems exhibit phenomena absent in Hermitian counterparts, many of which are associated with exceptional points (EPs)---non-Hermitian degeneracies at which both eigenvalues and eigenvectors coalesce and the spectral structure becomes singular. See Refs.~\cite{AshidaGongUeda2020,BergholtzBudichKunst2021,MiriAlu2019,DingFangMa2022,OkumaSato2023} for reviews.

Parity-time (PT) symmetry plays a central role in exploring EP physics~\cite{BenderBoettcher1998,Bender2007, ElGanainyMakrisKhajavikhan2018,OzdemirRotterNoriYang2019,BenderHook2024}. In PT-symmetric systems, the spectrum can remain entirely real in an unbroken phase despite the lack of Hermiticity, while a transition to a PT-broken phase occurs through EPs where real eigenvalues merge and become complex. In this sense, EPs often serve as the critical points of real-to-complex spectral transitions and provide a natural starting point for understanding non-Hermitian quantum criticality~\cite{AshidaFurukawaUeda2017,KawabataAshidaUeda2017}.

On the other hand, conformal invariance is expected to emerge at critical systems where the correlation length diverges, as is typically the case for 1+1-dimensional gapless Hermitian systems, and conformal field theory (CFT) provides an effective description of such systems. When Hermiticity is relaxed, however, it is unclear whether gaplessness alone still guarantees conformality and whether new universal structures beyond the conventional CFT paradigm are required. Addressing these questions is particularly important in view of the growing number of non-Hermitian systems whose critical points are governed by EPs.

Indeed, recent studies of PT-symmetric lattice models have reported unusual universal signatures at EP criticality, such as entanglement-entropy scaling consistent with a negative or even complex-valued central charge~\cite{CouvreurJacobsenSaleur2017, ChangYouWenRyu2020,Lee2022,TuTzengChang2022, FossatiAresCalabrese2023,YangLuLu2024,ShimizuKawabata2025}. These observations strongly suggest an underlying non-unitary conformal structure, but leave open how such behavior emerges from microscopic lattice models and how it should be characterized in a controlled field-theoretic framework.

In this work, we introduce a PT-symmetric non-Hermitian free-fermion field theory in 1+1 dimensions, which admits an exact quantization within a biorthogonal formalism. We show that this continuum theory realizes a Virasoro algebra with a negative central charge and exhibits universal features closely connected to \textit{logarithmic conformal field theory (LCFT)}~\cite{Gurarie1993,Flohr2003,CreutzigRidout2013}. Building on the field-theory analysis, we then establish a concrete lattice-continuum matching for EP criticality in a non-Hermitian lattice model which provides a microscopic construction of the proposed field theory. By identifying the conformal symmetry on the lattice, we extract the same LCFT data from finite-size computations, demonstrating quantitative agreement between the continuum theory and the lattice description.

\textit{PT-symmetric free-fermion field theory in 1+1 dimensions---}
Consider a non-Hermitian free-fermion field theory in 1+1 dimensions on a spatial circle of circumference $L$:
\begin{align}
\label{PTsymm_action}
S = \int_{-\infty}^{\infty} dt\int_0^{L} dx &\left[i\psi_+^\dagger\left(\partial_t+v\partial_x\right)\psi_+
+i\psi_-^\dagger\left(\partial_t-v\partial_x\right)\psi_-
\right.\nonumber\\
&\left.-\Delta\psi_+^\dagger\psi_-
-u(\partial_x\psi_-^{\dagger})(\partial_x\psi_+)\right],
\end{align}
where $\psi_\pm(x,t)$ are single-component complex fermions and $v,\Delta,u$ are free parameters. 
The terms proportional to $\Delta$ and $u$ explicitly break Hermiticity. Nevertheless, the theory has a PT symmetry provided that $v$ is real and one of the following conditions holds:
\begin{align}
\label{PT_symm_field}
\text{(i)}\ \Delta, u&\in\mathbb{R},\quad \psi_{\pm}(x,t)\xrightarrow{PT} \psi_{\pm}(-x,-t),
\nonumber\\
\text{(ii)}\ \Delta, u&\in i\mathbb{R},\quad \psi_{\pm}(x,t)\xrightarrow{PT} \pm\psi_{\pm}(-x,-t).
\end{align}
When $u=0$, the theory can be represented as the non-Hermitian Dirac theory with equal (up to a sign) Dirac and $\gamma_5$ mass coefficients discussed in~\cite{BenderJonesRivers2005,Alexandre:2017foi,Dora:2021cxu}. Throughout this paper, we will mainly consider the PT-symmetric cases.

Because $S^\dagger\neq S$, the equations of motion for $\psi_\pm$ and $\psi^\dagger_\pm$ are not related by Hermitian conjugation and are therefore inequivalent. We distinguish them by writing $\psi^L_\pm$ for solutions of $\frac{\delta_L S}{\delta\psi^{\dagger}_{\pm}}=0$ and $(\psi^R_\pm)^\dagger$ for solutions of $\frac{\delta_R S}{\delta\psi_{\pm}}=0$, where $\delta_{L/R}$ denote left/right functional derivatives. Canonical quantization is then implemented by imposing the equal-time anti-commutation relations between biorthogonal---left and right---fields:
\begin{align}
&\left\{\psi^L_\alpha(x,t),[\psi^{R}_{\beta}(y,t)]^{\dagger}\right\}
=\delta_{\alpha\beta}\delta\left(x-y\right), \nonumber \\
&\left\{\psi^L_\alpha(x,t),\psi_\beta^R(y,t)\right\}=
\left\{[\psi^{L}_{\alpha}(x,t)]^{\dagger},[\psi^{R}_{\beta}(y,t)]^{\dagger}\right\}=0.
\end{align}
With a periodic boundary condition in space, both zero and nonzero fermionic modes are present. Diagonalizing the Hamiltonian via mode expansions of $\psi_\pm^L$ and $(\psi_\pm^R)^\dagger$ yields the total Hamiltonian and momentum~\cite{SM}:
\begin{align}
\label{Hamiltonian_momentum}
H&=\Delta(\psi_{+,0}^{R})^{\dagger}\psi_{-,0}^L\nonumber\\
&\quad\ +\frac{2\pi v_F}{L}\sum_{m\in\mathbb{Z}\backslash\{0\}}m\left[(B_m^R)^{\dagger}B_m^L-(C_m^R)^{\dagger}C_m^L\right], 
\nonumber\\
P&=\frac{2\pi}{L}\sum_{m\in\mathbb{Z}\backslash\{0\}}m\left[(B_m^R)^{\dagger}B_m^L+(C_m^R)^{\dagger}C_m^L\right], 
\end{align}
where the effective Fermi velocity is $v_F=\sqrt{v^2+u\Delta}$ and the mode operators satisfy the biorthogonal anti-commutation relations
$\{\psi_{\alpha,0}^L,(\psi_{\beta,0}^{R})^{\dagger}\}=\delta_{\alpha\beta}$ and
$\{B_m^L,(B_n^R)^{\dagger}\}=\{C_m^L,(C_n^R)^{\dagger}\}=\delta_{mn}$,
with all remaining anticommutators vanishing.
In the presence of PT symmetry, $u\Delta$ is real, and we require $v^2>-u\Delta$ so that $v_F$ is also real. The spectrum is gapless in the scaling limit $L\to\infty$ and exhibits linear dispersion in momentum. The appearance of such a linear spectrum in the presence of a second-order spatial derivative reflects an accidental relativistic structure of the theory, even though Lorentz symmetry is not manifest in the action.

The gaplessness of the system is somewhat counterintuitive from a naive RG viewpoint: $\Delta$ has positive mass dimension and would be regarded as relevant, whereas $u$ has negative mass dimension and seems irrelevant. In the present non-Hermitian free theory, however, these parameters enter low-energy observables through marginal combination $u\Delta$, e.g., through the renormalized velocity $v_F$.

There are four types of excitations identified from~\eqref{Hamiltonian_momentum}: right-moving particles and antiparticles created by $(B_m^R)^{\dagger}$ and $B_{-m}^L$, respectively, and left-moving particles and antiparticles created by $(C_{-m}^R)^{\dagger}$ and $C_m^L$,  respectively, all with $m>0$.
When $\Delta=0$, the spectrum is essentially that of a massless Dirac theory, despite the presence of the non-Hermitian term  $u\neq 0$. 
In this case the theory can be mapped to the Hermitian theory ($\Delta=u=0$) by a similarity transformation while preserving the spectrum. 
When $\Delta\neq 0$, the zero-mode sector becomes non-diagonalizable, suggesting a critical theory distinct from the Hermitian case.

\textit{Conformal invariance---}
Given the gapless spectrum with linear dispersion, it is natural to ask whether the theory realizes full two-dimensional conformal symmetry, as is generally the case for Hermitian gapless theories in 1+1 dimensions. A standard diagnostic is whether the theory admits a traceless energy-momentum tensor. The canonical energy-momentum tensor $T_c^{\mu\nu}$ (with $\mu,\nu=t,x$) derived from the Lagrangian density in~\eqref{PTsymm_action} has trace
\begin{align}
(T_c)^{\mu}{}_{\mu}=&-i\psi^\dagger_+\left(\partial_t+v\partial_x\right)\psi_+-i\psi^\dagger_-\left(\partial_t-v\partial_x\right)\psi_-
\nonumber\\
&+2\Delta\psi^\dagger_+\psi_-.
\end{align}
which does not vanish even after imposing the equations of motion, except when $\Delta=0$. Remarkably, $T_c^{\mu\nu}$ admits an improvement~\cite{SM}
\begin{align}
T^{\mu\nu}=T_c^{\mu\nu}+\partial_\alpha \varepsilon^{\mu\alpha} I^\nu ,
\end{align}
such that the improved energy-momentum tensor becomes traceless on shell. Here $\varepsilon^{\mu\alpha}$ is the Levi-Civita tensor and the current $I^\nu$ has components $I_0=-iv\,\psi_-^\dagger\psi_-$ and $I_1=-i\,\psi_-^\dagger\psi_-$. In the relativistic limit $u=0$, $I^\nu$ reduces (after rescaling by the velocity $v$) to the left-moving chiral current density, up to a factor of $i$.

The existence of an on-shell traceless energy-momentum tensor motivates an explicit construction of conformal generators from $T^{\mu\nu}$. We define the dimensionless higher Hamiltonians and momenta
\begin{align}
\label{HnPn_field}
H_{n} &:= \frac{L}{2\pi v_F}\int dx~e^{-i\frac{2\pi n}{L}x}T^{00} + \frac{c}{12}\delta_{n,0},
\nonumber\\
P_{n} &:= \frac{L}{2\pi}\int dx~e^{-i\frac{2\pi n}{L}x}T^{01},
\end{align}
where $c$ is the central charge to be determined, and relate them to holomorphic/antiholomorphic generators by
\begin{align}
\label{HnPntoLnbarL_n}
H_{n} = L_n + \bar{L}_{-n},
\quad
P_{n} = L_n - \bar{L}_{-n}.
\end{align}
Specifically, the holomorphic generator takes the form
\begin{align}
\label{Virasoro_generator}
L_{n} &=
\begin{cases}
(B^{R}_{-n})^{\dagger}\psi^{L}_{-,0}
+ n\,(\psi^{R}_{+,0})^{\dagger} B^{L}_{n}
\\[0.2em]
\quad
+ \displaystyle\sum_{m\neq 0,n}
  m\,(B^{R}_{m-n})^{\dagger} B^{L}_{m},
& n\neq 0,
\\[0.9em]
(\psi^{R}_{+,0})^{\dagger}\psi^{L}_{-,0}
\\[0.2em]
\quad
+ \displaystyle\sum_{m\neq 0}
  m\,(B^{R}_{m})^{\dagger} B^{L}_{m}
+ \frac{c}{24},
& n=0,
\end{cases}
\end{align}
with an analogous expression for $\bar{L}_{n}$ (involving $C^{R/L}_{m}$). Here $\psi_{\pm,0}^{R/L}$ (with a rescaled form) and $B^{R/L}_{m}$---and $C^{R/L}_{m}$ in $\bar{L}_{n}$---are the fermionic mode operators that diagonalize the Hamiltonian~\eqref{Hamiltonian_momentum}.
One can explicitly verify that $L_n$ and $\bar{L}_{n}$ satisfy the Virasoro algebra---algebra of two-dimensional local conformal symmetry:
\begin{align}
\label{Virasoro}
[L_m, L_n] &= (m - n)L_{m+n} + \frac{c}{12}m(m^2 - 1)\delta_{m+n, 0},
\nonumber\\
[\bar{L}_m, \bar{L}_n] &= (m - n)\bar{L}_{m+n} + \frac{\bar{c}}{12}m(m^2 - 1)\delta_{m+n, 0},
\nonumber\\
[L_m, \bar{L}_n] &= 0.
\end{align}
This construction yields a negative central charge $c=\bar{c}=-2$~\cite{SM}, reflecting the intrinsically non-Hermitian character of the theory when $\Delta\neq0$. This should be contrasted with the Hermitian limit $\Delta=u=0$---and more generally with its similarity-equivalent counterpart $\Delta=0$ with $u\neq 0$---in which the model reduces to the Dirac CFT with $c=1$.

\textit{Connection to logarithmic conformal field theory---}
Besides the negative central charge, the conformally invariant non-Hermitian field theory exhibits extra features that are also absent in unitary CFT.
Most notably, correlation functions can display logarithmic scaling (in addition to power-law scaling). Concretely, the biorthogonal equal-time two-point functions
\(
G^{RL}_{\alpha\beta}(x; y)=\langle 0|[\psi^R_\alpha(x,t)]^\dagger \psi^L_\beta(y,t)|0\rangle
\)
in the large $L$ limit have the following scaling forms~\cite{SM}:
\begin{align}
G^{RL}_{++}(x; y) &\sim \frac{v}{2\pi i v_f}\,\frac{1}{x-y}, \nonumber\\
G^{RL}_{--}(x; y) &\sim -\frac{v}{2\pi i v_f}\,\frac{1}{x-y}, \nonumber\\
G^{RL}_{+-}(x; y) &\sim \frac{u}{2\pi v_f}\,\frac{1}{(x-y)^2}, \nonumber\\
G^{RL}_{-+}(x; y) &\sim \frac{\Delta}{2\pi v_f}\,
\ln\!\left(\frac{|x-y|}{L}\right).
\label{eq:log-propagator-scaling}
\end{align}
The last one exhibits the characteristic logarithm.  While the overall coefficient in
\(G^{RL}_{-+}\) depends on the normalization of the local fields, the appearance of an additive logarithm is a robust manifestation of scale invariance in a theory with non-diagonalizable dilatation operators and is a standard hallmark of logarithmic conformal field theory~\cite{Gurarie1993,Flohr2003,CreutzigRidout2013}.

In an LCFT, the spectrum of the (holomorphic) dilatation operator \(L_0\) forms reducible yet indecomposable Virasoro representations---staggered modules.
To characterize this structure in our non-Hermitian theory, we again work in the biorthogonal formalism.
For an eigenstate \(|v^R\rangle\) of \(L_0\), its biorthogonal norm is \(\langle v^L|v^R\rangle\),
where \(|v^L\rangle\) is the corresponding eigenstate of \(L_0^\dagger\).
In this sense, two of the fourfold-degenerate ground states,
\(
|\phi_0^R\rangle=(\psi_{+,0}^R)^\dagger|0\rangle
\)
and
\(
|\psi_0^R\rangle=(\psi_{-,0}^R)^\dagger|0\rangle
\),
have vanishing norms and form a rank-$1$ Jordan block,
\begin{align}
L_0\,|\phi^{R}_0\rangle= 0,
\qquad
L_0\,|\psi^{R}_0\rangle = |\phi^{R}_0\rangle .
\end{align}
Such a pair \((|\psi^R\rangle,|\phi^R\rangle)\) is referred to as a logarithmic pair.
In what follows we focus on the holomorphic sector; the antiholomorphic sector is completely analogous.

The Virasoro staggered modules are built from higher logarithmic pairs $(|\psi^{R}\rangle,|\phi^{R}\rangle)$ together with a certain set of $L_0$ eigenstates $|\xi^{R}\rangle$, and one can associate each such module with an integer $M\ge 1$ (the level). These states can be constructed, up to normalization, from the mode operators in~\eqref{Virasoro_generator} as
\begin{align}
\label{eq:psi_phi_xi_modes_refined}
|\phi_M^R\rangle&=(B_M^R)^\dagger\cdots(B_1^R)^\dagger
(\psi_{+,0}^R)^\dagger|0\rangle,
\nonumber\\
|\psi_M^R\rangle&=(B_M^R)^\dagger\cdots(B_1^R)^\dagger
(\psi_{-,0}^R)^\dagger|0\rangle,
\nonumber\\
|\xi_M^R\rangle&=
\begin{cases}
(\psi_{+,0}^R)^\dagger(\psi_{-,0}^R)^\dagger|0\rangle,
\quad M=1,
\\[0.3em]
(B_{M-1}^R)^\dagger\cdots(B_1^R)^\dagger
(\psi_{+,0}^R)^\dagger(\psi_{-,0}^R)^\dagger|0\rangle,
\quad M>1.
\end{cases}
\end{align}
One finds that $|\phi^{R}_M\rangle$ and $|\xi^{R}_M\rangle$ are highest-weight eigenstates of $L_0$ with eigenvalues (conformal weights)
$h_M=\sum_{k=1}^{M}k=\frac{M(M+1)}{2}$ and
$h_{M-1}=\sum_{k=1}^{M-1}k=\frac{M(M-1)}{2}$, 
while $|\psi^{R}_M\rangle$ is a generalized eigenstate, as reflected in the Jordan action of $L_0$:
\begin{equation}
\label{eq:L0Jordan}
L_0|\psi^{R}_M\rangle = h_M|\psi^{R}_M\rangle + |\phi^{R}_M\rangle.
\end{equation}
Moreover, $|\phi^{R}_M\rangle$ and $|\psi_M^R\rangle$ are both singular (null) states with vanishing biorthogonal norm.

A defining feature of a staggered module is that the singular highest-weight state $|\phi^{R}_M\rangle$ also arises as a Virasoro descendant of $|\xi^{R}_M\rangle$:
\begin{equation}
\label{eq:phi_as_singular}
 A_{M,-}\,|\xi^{R}_M\rangle=a_{M}|\phi^{R}_M\rangle,
\end{equation}
where $A_{M,-}$ is a level-$M$ singular-vector operator composed of Virasoro generators and $a_M$ is a proportionality constant. In the normalization where the coefficient of $L_{-1}^M$ is $1$, $A_{M,-}$ is given by~\cite{BenoitSaintAubin1988}
\begin{align}
\label{eq:BSA}
A_{M,-}
=\sum_{\substack{\text{$(k_1,k_2,\ldots,k_r)\in$}\\ \text{compositions of $M$}}}
(-2)^{\,M-r}\,C(k_{1},\ldots,k_{r})\,
L_{-k_{1}}\cdots L_{-k_{r}},
\end{align}
where $C(k_1,\ldots,k_r)$ is the inverse of
\begin{align}
[(M-1)!]^2\prod_{i=1}^{r-1}\Big[(k_1+\cdots+k_i)(k_{i+1}+\cdots+k_r)\Big].
\nonumber
\end{align}

\begin{figure}[t]
    \centering
    \includegraphics[scale=0.6]{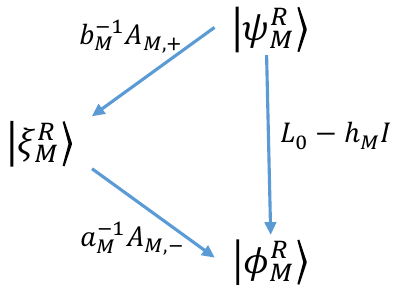}
    \caption{Level-$M$ holomorphic staggered module.}
    \label{fig:staggered_module}
\end{figure}


\begin{table}[h]
\centering
\begin{tabular}{c|c}
\hline\hline
$M$ & $\beta_M$ \\
\hline
1 &  $-1$ \\
2 &  $-18$ \\
3 &  $-2700$ \\
4 &  $-1587600$ \\
5 &  $-2571912000$ \\
\hline\hline
\end{tabular}
\caption{Explicit values of indecomposability parameters up to level five.}
\label{tab:abeta_levels}
\end{table}

To quantify the extension data of this staggered module, we also need the relation between $|\xi^{R}_M\rangle$ and $|\psi^{R}_M\rangle$. They are related by the biorthogonal adjoint of $A_{M,-}$, denoted $A_{M,+}$, which takes the form
\begin{equation}
A_{M,+}
=\sum_{\substack{\text{$(k_1,k_2,\ldots,k_r)\in$}\\ \text{compositions of $M$}}}
(-2)^{\,M-r}\,C(k_{1},\ldots,k_{r})\,
L_{k_{r}}\cdots L_{k_{1}},
\end{equation}
with the same coefficients $C(\cdot)$ as in~\eqref{eq:BSA}. Specifically,
\begin{equation}
\label{eq:beta_def}
A_{M,+}\,|\psi^{R}_M\rangle=b_{M}\,|\xi^{R}_M\rangle,
\end{equation}
defining a second constant $b_M$. This completes the level-$M$ staggered module shown in Fig.~\ref{fig:staggered_module}. A characteristic number of this module, which we call the level-$M$ indecomposability parameter, is defined as
\begin{equation}
\beta_M=a_{M}b_{M}.
\end{equation}
Note that $\beta_M$ depends only on the normalization of $A_{M,\pm}$, not on that of the primary state $|\xi^{R}_M\rangle$. 
In the standard LCFT framework, $\beta_M$ is an intrinsic invariant of the staggered module that measures the nontrivial extension of a conventional CFT  through its Jordan-cell structure.
Using the explicit form of $L_n$ in~\eqref{Virasoro_generator}, one obtains~\cite{SM}
\begin{equation}
\label{eq:beta_formula}
\beta_M = -\,\frac{M\big[(2M-1)!\big]^2}{4^{\,M-1}}.
\end{equation}
The values of $\beta_M$ up to $M=5$ are listed in Table~\ref{tab:abeta_levels}.
The antiholomorphic sector admits an entirely analogous construction and yields the same indecomposability parameters. 

The formula~\eqref{eq:beta_formula} coincides with the characteristic indecomposability parameters in the $c=-2$ symplectic fermion theory~\cite{Kausch1995, Kausch2000, Ryu:2022ffg}, where they can be extracted, e.g., from fusion products of indecomposable Virasoro representations~\cite{GaberdielKausch1996} and have been further discussed in subsequent works~\cite{VasseurJacobsenSaleur2011, VasseurThesis2013}.
However, we emphasize that the two theories are not identical. Symplectic fermions are built from anticommuting scalar fields, whereas the present model is a non-Hermitian theory of conventional (spinor) fermions. In our case, the logarithmic structure originates from non-diagonalizability induced by the non-Hermitian structure and is naturally formulated within a biorthogonal Hilbert space. Thus, while the Virasoro staggered-module data agree at the level of representation theory, the spectrum of local operators and the extended chiral algebra are expected to differ between the two realizations.

\textit{Lattice realization---}
We now present a lattice realization of the field theory~\eqref{PTsymm_action} and explore its LCFT structure at exceptional-point criticality. Consider a one-dimensional tight-binding model with alternating hopping amplitudes and a staggered onsite potential:
$H_{\mathrm{TB}}=\sum_{j=1}^{N}h_j$ (with $N$ even and periodic boundary conditions), where
\begin{align}
h_j=-[t+(-1)^j\delta]\,(c^\dagger_{j+1}c_j+c^\dagger_j c_{j+1})-(-1)^j\mu_s\,c^\dagger_j c_j.
\label{latt_hj}
\end{align}
This model admits two inequivalent implementations of parity, corresponding to site- and bond-centered reflections, and hence two distinct realizations of PT symmetry. In the site-centered case, PT symmetry requires $t,\mu_s\in\mathbb{R}$ and $\delta\in i\mathbb{R}$, which includes the non-Hermitian model discussed in~\cite{AshidaUeda2018, Dora:2021cxu, MedenGrunwaldKennes2023}. In the bond-centered case, PT symmetry requires $t,\delta\in\mathbb{R}$ and $\mu_s\in i\mathbb{R}$, as in the non-Hermitian SSH model studied in~\cite{Lieu2018, ChangYouWenRyu2020}. In the continuum limit, these two lattice realizations correspond to the two PT actions in~\eqref{PT_symm_field}.

The single-particle gap is proportional to $\sqrt{\mu_s^2+4\delta^2}$ and closes at
$\mu_s=\pm 2i\delta$, where the Bloch Hamiltonian becomes defective and realizes a second-order EP.
The resulting critical theory has a linear low-energy dispersion with Fermi velocity
$v_F=2\sqrt{t^2-\delta^2}$.
Expanding around the EP $\mu_s=2i\delta$ yields the continuum theory~\eqref{PTsymm_action} with the parameter
identification 
\begin{align}
v=2t, \qquad\Delta=4i\delta, \qquad u=i\delta,
\end{align}
while the EP manifests itself as a Jordan structure in the low-energy sector.
Guided by the previous field-theory analysis, we expect that the EP critical lattice model realizes a
$c=-2$ LCFT. Additional evidence comes from the non-Hermitian SSH chain ($t,\delta\in\mathbb{R}$ and
$\mu_s\in i\mathbb{R}$), whose EP critical point exhibits logarithmic entanglement-entropy scaling
consistent with $c=-2$~\cite{ChangYouWenRyu2020,Lee2022,TuTzengChang2022,YangLuLu2024}.

Before identifying the full conformal structure, we first compute the lattice correlation function in the
biorthogonal formalism.  In the large $N$ limit, we find~\cite{SM}
\begin{align}
\langle c^\dagger_j c_{j'}\rangle
\;\sim\;
(-1)^j\, i^{\,j-j'}\,\kappa\,
\ln\!\left(\frac{|j-j'|}{N}\right),
\label{eq:latt-log-Green}
\end{align}
where $\kappa \propto \frac{\delta}{v_F}$ is a nonuniversal amplitude.
This logarithmic scaling agrees with the continuum prediction~\eqref{eq:log-propagator-scaling},
up to the expected oscillatory phase factors dictated by the lattice-continuum embedding.

To test conformal symmetry directly on the lattice, we employ the Koo-Saleur construction~\cite{KooSaleur1994,MilstedVidal2017}.
Starting from the local Hamiltonian density $h_j$, we define the lattice momentum density via locality
and energy-momentum conservation,
\begin{align}
p_j := i[h_j,h_{j+1}].
\end{align}
We then introduce Fourier modes of the two densities,
\begin{align}
H^{\mathrm{latt}}_n
&=L^{\mathrm{latt}}_n+\bar L^{\mathrm{latt}}_{-n}
:=\frac{Na}{2\pi v_F}\sum_{j=1}^N e^{-i\frac{2\pi n}{N}j}\,(h_j-h_\infty)
+\frac{c}{12}\delta_{n,0},
\nonumber\\
P^{\mathrm{latt}}_n
&=L^{\mathrm{latt}}_n-\bar L^{\mathrm{latt}}_{-n}
:=\frac{Na^2}{2\pi v_F^2}\sum_{j=1}^N e^{-i\frac{2\pi n}{N}j}\,p_j,
\end{align}
which are lattice analogs of~\eqref{HnPn_field}.  
Here $a$ is the lattice spacing and $h_\infty$ denotes the ground-state energy density in the thermodynamic limit.
As in the continuum, where the canonical energy-momentum tensor  needs to be improved by total-derivative terms to realize
conformal symmetry, the Virasoro algebra is not obtained directly from $L_n^{\mathrm{latt}}$ and
$\bar L_n^{\mathrm{latt}}$ constructed from the bare density~\eqref{latt_hj}.  Instead, it is
restored only after shifting the local density by a lattice total derivative.  For example, at
$\mu_s=2i\delta$ we implement
\begin{align}
h_j \;\rightarrow\; &\,h_j
-it\left(c^\dagger_{j}c_{j}-c^\dagger_{j+1}c_{j+1}\right)
\nonumber\\
&\,+(-1)^{j}i\delta\left(c^\dagger_{j}c_{j}+c^\dagger_{j+1}c_{j+1}\right),
\end{align}
which leaves the total Hamiltonian and momentum unchanged while modifying their higher modes.
In the continuum limit, this modification reproduces the improved traceless energy-momentum tensor  of the field theory.
As expected, the resulting $L_n^{\mathrm{latt}}$ and $\bar L_n^{\mathrm{latt}}$ converge to the
Virasoro algebra with $c=-2$ in the scaling limit, consistent with entanglement-based diagnostics.

Using the modified lattice Virasoro operators, one can construct staggered modules and extract indecomposability parameters from finite-size spectra and the associated eigenstates. For instance, for the level-1 and level-2 modules we obtain~\cite{SM}
\begin{align}
\beta^{\mathrm{latt}}_1&=-1+\frac{17}{48}\left(\frac{2\pi}{N}\right)^2+\mathcal{O}(N^{-4}),
\nonumber\\
\beta^{\mathrm{latt}}_2&=-18+\frac{125}{4}\left(\frac{2\pi}{N}\right)^2+\mathcal{O}(N^{-4}),
\end{align}
in agreement with the field-theory result~\eqref{eq:beta_formula}.

\textit{Conclusion and outlook---}
Our results sharpen the notion of universality without Hermiticity. In this work, we show that conformal invariance can emerge in a genuinely non-Hermitian gapless system. For a PT-symmetric free-fermion theory with linear dispersion and an exceptional-point singularity, we make the conformal structure explicit within the biorthogonal framework. A key message is that the standard link between a traceless energy-momentum tensor and conformal invariance~\cite{DiFrancescoMathieuSenechal1997, BelavinPolyakovZamolodchikov1984, Polchinski1988, Nakayama2015}
extends to non-Hermitian theories once locality and dynamics are formulated consistently in this setting. On the lattice, the same principle becomes concrete. Conformal symmetry is recovered only after an improvement (associated with a lattice total derivative) of the local Hamiltonian density, which in turn allows the conformal structure to be diagnosed directly from the lattice model.

The resulting conformal data are ``unconventional" from the viewpoint of unitary CFTs: the central charge is negative, correlators display logarithmic scaling, and the spectrum forms indecomposable Virasoro representations. 
Remarkably, the associated indecomposability parameters coincide with those of the $c=-2$ symplectic fermion LCFT despite distinct microscopic realizations and local operator content, indicating that logarithmic extension data provide robust universal invariants.

Natural future directions include moving beyond free-field theories by introducing controlled interaction deformations and developing renormalization-group analyses of interacting non-Hermitian theories, aided by dual formulations that allow universal data to be tracked across fermionic and bosonic descriptions. 
It will also be interesting to incorporate boundaries and impurities, where non-Hermitian boundary field theories and their logarithmic extensions can provide a systematic description of boundary criticality and impurity responses. 
Connecting these developments to measurable consequences via physical response functions and experimentally relevant observables---including transport properties, dynamical signatures, and entanglement measures---should enable decisive validation of the theoretical predictions.



\begin{acknowledgements}
\textit{Acknowledgements---}
The authors thank Po-Yao Chang, Ken Kikuchi, Yuan Miao, Jiaxin Qiao, Shinsei Ryu, and Yuji Tachikawa for valuable discussions.
C.-T. H. is supported by the Yushan (Young) Scholar Program under Grant No. NTU-111VV016 and by the National Science and Technology Council (NSTC) of Taiwan under Grant No. 112-2112-M-002-048-MY3.
\end{acknowledgements}

\bibliography{bib}

\begin{thebibliography}{42}%
\makeatletter
\providecommand \@ifxundefined [1]{%
 \@ifx{#1\undefined}
}%
\providecommand \@ifnum [1]{%
 \ifnum #1\expandafter \@firstoftwo
 \else \expandafter \@secondoftwo
 \fi
}%
\providecommand \@ifx [1]{%
 \ifx #1\expandafter \@firstoftwo
 \else \expandafter \@secondoftwo
 \fi
}%
\providecommand \natexlab [1]{#1}%
\providecommand \enquote  [1]{``#1''}%
\providecommand \bibnamefont  [1]{#1}%
\providecommand \bibfnamefont [1]{#1}%
\providecommand \citenamefont [1]{#1}%
\providecommand \href@noop [0]{\@secondoftwo}%
\providecommand \href [0]{\begingroup \@sanitize@url \@href}%
\providecommand \@href[1]{\@@startlink{#1}\@@href}%
\providecommand \@@href[1]{\endgroup#1\@@endlink}%
\providecommand \@sanitize@url [0]{\catcode `\\12\catcode `\$12\catcode
  `\&12\catcode `\#12\catcode `\^12\catcode `\_12\catcode `\%12\relax}%
\providecommand \@@startlink[1]{}%
\providecommand \@@endlink[0]{}%
\providecommand \url  [0]{\begingroup\@sanitize@url \@url }%
\providecommand \@url [1]{\endgroup\@href {#1}{\urlprefix }}%
\providecommand \urlprefix  [0]{URL }%
\providecommand \Eprint [0]{\href }%
\providecommand \doibase [0]{http://dx.doi.org/}%
\providecommand \selectlanguage [0]{\@gobble}%
\providecommand \bibinfo  [0]{\@secondoftwo}%
\providecommand \bibfield  [0]{\@secondoftwo}%
\providecommand \translation [1]{[#1]}%
\providecommand \BibitemOpen [0]{}%
\providecommand \bibitemStop [0]{}%
\providecommand \bibitemNoStop [0]{.\EOS\space}%
\providecommand \EOS [0]{\spacefactor3000\relax}%
\providecommand \BibitemShut  [1]{\csname bibitem#1\endcsname}%
\let\auto@bib@innerbib\@empty
\bibitem [{\citenamefont {Ashida}\ \emph {et~al.}(2020)\citenamefont {Ashida},
  \citenamefont {Gong},\ and\ \citenamefont {Ueda}}]{AshidaGongUeda2020}%
  \BibitemOpen
  \bibfield  {author} {\bibinfo {author} {\bibfnamefont {Y.}~\bibnamefont
  {Ashida}}, \bibinfo {author} {\bibfnamefont {Z.}~\bibnamefont {Gong}}, \ and\
  \bibinfo {author} {\bibfnamefont {M.}~\bibnamefont {Ueda}},\ }\href {\doibase
  10.1080/00018732.2021.1876991} {\bibfield  {journal} {\bibinfo  {journal}
  {Adv. Phys.}\ }\textbf {\bibinfo {volume} {69}},\ \bibinfo {pages} {249}
  (\bibinfo {year} {2020})},\ \Eprint {http://arxiv.org/abs/2006.01837}
  {arXiv:2006.01837 [cond-mat.mes-hall]} \BibitemShut {NoStop}%
\bibitem [{\citenamefont {Bergholtz}\ \emph {et~al.}(2021)\citenamefont
  {Bergholtz}, \citenamefont {Budich},\ and\ \citenamefont
  {Kunst}}]{BergholtzBudichKunst2021}%
  \BibitemOpen
  \bibfield  {author} {\bibinfo {author} {\bibfnamefont {E.~J.}\ \bibnamefont
  {Bergholtz}}, \bibinfo {author} {\bibfnamefont {J.~C.}\ \bibnamefont
  {Budich}}, \ and\ \bibinfo {author} {\bibfnamefont {F.~K.}\ \bibnamefont
  {Kunst}},\ }\href {\doibase 10.1103/RevModPhys.93.015005} {\bibfield
  {journal} {\bibinfo  {journal} {Rev. Mod. Phys.}\ }\textbf {\bibinfo {volume}
  {93}},\ \bibinfo {pages} {015005} (\bibinfo {year} {2021})},\ \Eprint
  {http://arxiv.org/abs/1912.10048} {arXiv:1912.10048 [cond-mat.mes-hall]}
  \BibitemShut {NoStop}%
\bibitem [{\citenamefont {Miri}\ and\ \citenamefont
  {Al{\`u}}(2019)}]{MiriAlu2019}%
  \BibitemOpen
  \bibfield  {author} {\bibinfo {author} {\bibfnamefont {M.-A.}\ \bibnamefont
  {Miri}}\ and\ \bibinfo {author} {\bibfnamefont {A.}~\bibnamefont {Al{\`u}}},\
  }\href {\doibase 10.1126/science.aar7709} {\bibfield  {journal} {\bibinfo
  {journal} {Science}\ }\textbf {\bibinfo {volume} {363}},\ \bibinfo {pages}
  {eaar7709} (\bibinfo {year} {2019})}\BibitemShut {NoStop}%
\bibitem [{\citenamefont {Ding}\ \emph {et~al.}(2022)\citenamefont {Ding},
  \citenamefont {Fang},\ and\ \citenamefont {Ma}}]{DingFangMa2022}%
  \BibitemOpen
  \bibfield  {author} {\bibinfo {author} {\bibfnamefont {K.}~\bibnamefont
  {Ding}}, \bibinfo {author} {\bibfnamefont {C.}~\bibnamefont {Fang}}, \ and\
  \bibinfo {author} {\bibfnamefont {G.}~\bibnamefont {Ma}},\ }\href {\doibase
  10.1038/s42254-022-00516-5} {\bibfield  {journal} {\bibinfo  {journal} {Nat.
  Rev. Phys.}\ }\textbf {\bibinfo {volume} {4}},\ \bibinfo {pages} {745}
  (\bibinfo {year} {2022})},\ \Eprint {http://arxiv.org/abs/2204.11601}
  {arXiv:2204.11601 [quant-ph]} \BibitemShut {NoStop}%
\bibitem [{\citenamefont {Okuma}\ and\ \citenamefont
  {Sato}(2023)}]{OkumaSato2023}%
  \BibitemOpen
  \bibfield  {author} {\bibinfo {author} {\bibfnamefont {N.}~\bibnamefont
  {Okuma}}\ and\ \bibinfo {author} {\bibfnamefont {M.}~\bibnamefont {Sato}},\
  }\href {\doibase 10.1146/annurev-conmatphys-040521-033133} {\bibfield
  {journal} {\bibinfo  {journal} {Annu. Rev. Condens. Matter Phys.}\ }\textbf
  {\bibinfo {volume} {14}},\ \bibinfo {pages} {83} (\bibinfo {year} {2023})},\
  \Eprint {http://arxiv.org/abs/2205.10379} {arXiv:2205.10379
  [cond-mat.mes-hall]} \BibitemShut {NoStop}%
\bibitem [{\citenamefont {Bender}\ and\ \citenamefont
  {Boettcher}(1998)}]{BenderBoettcher1998}%
  \BibitemOpen
  \bibfield  {author} {\bibinfo {author} {\bibfnamefont {C.~M.}\ \bibnamefont
  {Bender}}\ and\ \bibinfo {author} {\bibfnamefont {S.}~\bibnamefont
  {Boettcher}},\ }\href {\doibase 10.1103/PhysRevLett.80.5243} {\bibfield
  {journal} {\bibinfo  {journal} {Phys. Rev. Lett.}\ }\textbf {\bibinfo
  {volume} {80}},\ \bibinfo {pages} {5243} (\bibinfo {year} {1998})},\ \Eprint
  {http://arxiv.org/abs/physics/9712001} {arXiv:physics/9712001 [math-ph]}
  \BibitemShut {NoStop}%
\bibitem [{\citenamefont {Bender}(2007)}]{Bender2007}%
  \BibitemOpen
  \bibfield  {author} {\bibinfo {author} {\bibfnamefont {C.~M.}\ \bibnamefont
  {Bender}},\ }\href {\doibase 10.1088/0034-4885/70/6/R03} {\bibfield
  {journal} {\bibinfo  {journal} {Rep. Prog. Phys.}\ }\textbf {\bibinfo
  {volume} {70}},\ \bibinfo {pages} {947} (\bibinfo {year} {2007})},\ \Eprint
  {http://arxiv.org/abs/hep-th/0703096} {arXiv:hep-th/0703096 [hep-th]}
  \BibitemShut {NoStop}%
\bibitem [{\citenamefont {El-Ganainy}\ \emph {et~al.}(2018)\citenamefont
  {El-Ganainy}, \citenamefont {Makris}, \citenamefont {Khajavikhan},
  \citenamefont {Musslimani}, \citenamefont {Rotter},\ and\ \citenamefont
  {Christodoulides}}]{ElGanainyMakrisKhajavikhan2018}%
  \BibitemOpen
  \bibfield  {author} {\bibinfo {author} {\bibfnamefont {R.}~\bibnamefont
  {El-Ganainy}}, \bibinfo {author} {\bibfnamefont {K.~G.}\ \bibnamefont
  {Makris}}, \bibinfo {author} {\bibfnamefont {M.}~\bibnamefont {Khajavikhan}},
  \bibinfo {author} {\bibfnamefont {Z.~H.}\ \bibnamefont {Musslimani}},
  \bibinfo {author} {\bibfnamefont {S.}~\bibnamefont {Rotter}}, \ and\ \bibinfo
  {author} {\bibfnamefont {D.~N.}\ \bibnamefont {Christodoulides}},\ }\href
  {\doibase 10.1038/nphys4323} {\bibfield  {journal} {\bibinfo  {journal}
  {Nature Physics}\ }\textbf {\bibinfo {volume} {14}},\ \bibinfo {pages} {11}
  (\bibinfo {year} {2018})}\BibitemShut {NoStop}%
\bibitem [{\citenamefont {{\"O}zdemir}\ \emph {et~al.}(2019)\citenamefont
  {{\"O}zdemir}, \citenamefont {Rotter}, \citenamefont {Nori},\ and\
  \citenamefont {Yang}}]{OzdemirRotterNoriYang2019}%
  \BibitemOpen
  \bibfield  {author} {\bibinfo {author} {\bibfnamefont {{\c{S}}.~K.}\
  \bibnamefont {{\"O}zdemir}}, \bibinfo {author} {\bibfnamefont
  {S.}~\bibnamefont {Rotter}}, \bibinfo {author} {\bibfnamefont
  {F.}~\bibnamefont {Nori}}, \ and\ \bibinfo {author} {\bibfnamefont
  {L.}~\bibnamefont {Yang}},\ }\href {\doibase 10.1038/s41563-019-0304-9}
  {\bibfield  {journal} {\bibinfo  {journal} {Nature Materials}\ }\textbf
  {\bibinfo {volume} {18}},\ \bibinfo {pages} {783} (\bibinfo {year}
  {2019})}\BibitemShut {NoStop}%
\bibitem [{\citenamefont {Bender}\ and\ \citenamefont
  {Hook}(2024)}]{BenderHook2024}%
  \BibitemOpen
  \bibfield  {author} {\bibinfo {author} {\bibfnamefont {C.~M.}\ \bibnamefont
  {Bender}}\ and\ \bibinfo {author} {\bibfnamefont {D.~W.}\ \bibnamefont
  {Hook}},\ }\href {\doibase 10.1103/RevModPhys.96.045002} {\bibfield
  {journal} {\bibinfo  {journal} {Rev. Mod. Phys.}\ }\textbf {\bibinfo {volume}
  {96}},\ \bibinfo {pages} {045002} (\bibinfo {year} {2024})},\ \Eprint
  {http://arxiv.org/abs/2312.17386} {arXiv:2312.17386 [quant-ph]} \BibitemShut
  {NoStop}%
\bibitem [{\citenamefont {Ashida}\ \emph {et~al.}(2017)\citenamefont {Ashida},
  \citenamefont {Furukawa},\ and\ \citenamefont
  {Ueda}}]{AshidaFurukawaUeda2017}%
  \BibitemOpen
  \bibfield  {author} {\bibinfo {author} {\bibfnamefont {Y.}~\bibnamefont
  {Ashida}}, \bibinfo {author} {\bibfnamefont {S.}~\bibnamefont {Furukawa}}, \
  and\ \bibinfo {author} {\bibfnamefont {M.}~\bibnamefont {Ueda}},\ }\href
  {\doibase 10.1038/ncomms15791} {\bibfield  {journal} {\bibinfo  {journal}
  {Nat. Commun.}\ }\textbf {\bibinfo {volume} {8}},\ \bibinfo {pages} {15791}
  (\bibinfo {year} {2017})},\ \Eprint {http://arxiv.org/abs/1611.00396}
  {arXiv:1611.00396 [cond-mat.stat-mech]} \BibitemShut {NoStop}%
\bibitem [{\citenamefont {Kawabata}\ \emph {et~al.}(2017)\citenamefont
  {Kawabata}, \citenamefont {Ashida},\ and\ \citenamefont
  {Ueda}}]{KawabataAshidaUeda2017}%
  \BibitemOpen
  \bibfield  {author} {\bibinfo {author} {\bibfnamefont {K.}~\bibnamefont
  {Kawabata}}, \bibinfo {author} {\bibfnamefont {Y.}~\bibnamefont {Ashida}}, \
  and\ \bibinfo {author} {\bibfnamefont {M.}~\bibnamefont {Ueda}},\ }\href
  {\doibase 10.1103/PhysRevLett.119.190401} {\bibfield  {journal} {\bibinfo
  {journal} {Phys. Rev. Lett.}\ }\textbf {\bibinfo {volume} {119}},\ \bibinfo
  {pages} {190401} (\bibinfo {year} {2017})},\ \Eprint
  {http://arxiv.org/abs/1705.04628} {arXiv:1705.04628 [quant-ph]} \BibitemShut
  {NoStop}%
\bibitem [{\citenamefont {Couvreur}\ \emph {et~al.}(2017)\citenamefont
  {Couvreur}, \citenamefont {Jacobsen},\ and\ \citenamefont
  {Saleur}}]{CouvreurJacobsenSaleur2017}%
  \BibitemOpen
  \bibfield  {author} {\bibinfo {author} {\bibfnamefont {R.}~\bibnamefont
  {Couvreur}}, \bibinfo {author} {\bibfnamefont {J.~L.}\ \bibnamefont
  {Jacobsen}}, \ and\ \bibinfo {author} {\bibfnamefont {H.}~\bibnamefont
  {Saleur}},\ }\href {\doibase 10.1103/PhysRevLett.119.040601} {\bibfield
  {journal} {\bibinfo  {journal} {Phys. Rev. Lett.}\ }\textbf {\bibinfo
  {volume} {119}},\ \bibinfo {pages} {040601} (\bibinfo {year} {2017})},\
  \Eprint {http://arxiv.org/abs/1611.08506} {arXiv:1611.08506
  [cond-mat.stat-mech]} \BibitemShut {NoStop}%
\bibitem [{\citenamefont {Chang}\ \emph {et~al.}(2020)\citenamefont {Chang},
  \citenamefont {You}, \citenamefont {Wen},\ and\ \citenamefont
  {Ryu}}]{ChangYouWenRyu2020}%
  \BibitemOpen
  \bibfield  {author} {\bibinfo {author} {\bibfnamefont {P.-Y.}\ \bibnamefont
  {Chang}}, \bibinfo {author} {\bibfnamefont {J.-S.}\ \bibnamefont {You}},
  \bibinfo {author} {\bibfnamefont {X.}~\bibnamefont {Wen}}, \ and\ \bibinfo
  {author} {\bibfnamefont {S.}~\bibnamefont {Ryu}},\ }\href {\doibase
  10.1103/PhysRevResearch.2.033069} {\bibfield  {journal} {\bibinfo  {journal}
  {Phys. Rev. Research}\ }\textbf {\bibinfo {volume} {2}},\ \bibinfo {pages}
  {033069} (\bibinfo {year} {2020})},\ \Eprint
  {http://arxiv.org/abs/1909.01346} {arXiv:1909.01346 [cond-mat.str-el]}
  \BibitemShut {NoStop}%
\bibitem [{\citenamefont {Lee}(2022)}]{Lee2022}%
  \BibitemOpen
  \bibfield  {author} {\bibinfo {author} {\bibfnamefont {C.~H.}\ \bibnamefont
  {Lee}},\ }\href {\doibase 10.1103/PhysRevLett.128.010402} {\bibfield
  {journal} {\bibinfo  {journal} {Phys. Rev. Lett.}\ }\textbf {\bibinfo
  {volume} {128}},\ \bibinfo {pages} {010402} (\bibinfo {year} {2022})},\
  \Eprint {http://arxiv.org/abs/2011.09505} {arXiv:2011.09505
  [cond-mat.quant-gas]} \BibitemShut {NoStop}%
\bibitem [{\citenamefont {Tu}\ \emph {et~al.}(2022)\citenamefont {Tu},
  \citenamefont {Tzeng},\ and\ \citenamefont {Chang}}]{TuTzengChang2022}%
  \BibitemOpen
  \bibfield  {author} {\bibinfo {author} {\bibfnamefont {Y.-T.}\ \bibnamefont
  {Tu}}, \bibinfo {author} {\bibfnamefont {Y.-C.}\ \bibnamefont {Tzeng}}, \
  and\ \bibinfo {author} {\bibfnamefont {P.-Y.}\ \bibnamefont {Chang}},\ }\href
  {\doibase 10.21468/SciPostPhys.12.6.194} {\bibfield  {journal} {\bibinfo
  {journal} {SciPost Physics}\ }\textbf {\bibinfo {volume} {12}},\ \bibinfo
  {pages} {194} (\bibinfo {year} {2022})},\ \Eprint
  {http://arxiv.org/abs/2107.13006} {arXiv:2107.13006 [cond-mat.str-el]}
  \BibitemShut {NoStop}%
\bibitem [{\citenamefont {Fossati}\ \emph {et~al.}(2023)\citenamefont
  {Fossati}, \citenamefont {Ares},\ and\ \citenamefont
  {Calabrese}}]{FossatiAresCalabrese2023}%
  \BibitemOpen
  \bibfield  {author} {\bibinfo {author} {\bibfnamefont {M.}~\bibnamefont
  {Fossati}}, \bibinfo {author} {\bibfnamefont {F.}~\bibnamefont {Ares}}, \
  and\ \bibinfo {author} {\bibfnamefont {P.}~\bibnamefont {Calabrese}},\ }\href
  {\doibase 10.1103/PhysRevB.107.205153} {\bibfield  {journal} {\bibinfo
  {journal} {Phys. Rev. B}\ }\textbf {\bibinfo {volume} {107}},\ \bibinfo
  {pages} {205153} (\bibinfo {year} {2023})},\ \Eprint
  {http://arxiv.org/abs/2303.05232} {arXiv:2303.05232 [cond-mat.stat-mech]}
  \BibitemShut {NoStop}%
\bibitem [{\citenamefont {Yang}\ \emph {et~al.}(2024)\citenamefont {Yang},
  \citenamefont {Lu},\ and\ \citenamefont {Lu}}]{YangLuLu2024}%
  \BibitemOpen
  \bibfield  {author} {\bibinfo {author} {\bibfnamefont {Z.}~\bibnamefont
  {Yang}}, \bibinfo {author} {\bibfnamefont {C.}~\bibnamefont {Lu}}, \ and\
  \bibinfo {author} {\bibfnamefont {X.}~\bibnamefont {Lu}},\ }\href {\doibase
  10.1103/PhysRevB.110.235127} {\bibfield  {journal} {\bibinfo  {journal}
  {Physical Review B}\ }\textbf {\bibinfo {volume} {110}},\ \bibinfo {pages}
  {235127} (\bibinfo {year} {2024})},\ \Eprint
  {http://arxiv.org/abs/2406.15564} {arXiv:2406.15564 [cond-mat.mes-hall]}
  \BibitemShut {NoStop}%
\bibitem [{\citenamefont {Shimizu}\ and\ \citenamefont
  {Kawabata}(2025)}]{ShimizuKawabata2025}%
  \BibitemOpen
  \bibfield  {author} {\bibinfo {author} {\bibfnamefont {H.}~\bibnamefont
  {Shimizu}}\ and\ \bibinfo {author} {\bibfnamefont {K.}~\bibnamefont
  {Kawabata}},\ }\href {\doibase 10.1103/n578-ljd5} {\bibfield  {journal}
  {\bibinfo  {journal} {Phys. Rev. B}\ }\textbf {\bibinfo {volume} {112}},\
  \bibinfo {pages} {085112} (\bibinfo {year} {2025})},\ \Eprint
  {http://arxiv.org/abs/2502.02001} {arXiv:2502.02001 [cond-mat.stat-mech]}
  \BibitemShut {NoStop}%
\bibitem [{\citenamefont {Gurarie}(1993)}]{Gurarie1993}%
  \BibitemOpen
  \bibfield  {author} {\bibinfo {author} {\bibfnamefont {V.}~\bibnamefont
  {Gurarie}},\ }\href {\doibase 10.1016/0550-3213(93)90528-W} {\bibfield
  {journal} {\bibinfo  {journal} {Nuclear Physics B}\ }\textbf {\bibinfo
  {volume} {410}},\ \bibinfo {pages} {535} (\bibinfo {year} {1993})},\ \Eprint
  {http://arxiv.org/abs/hep-th/9303160} {arXiv:hep-th/9303160} \BibitemShut
  {NoStop}%
\bibitem [{\citenamefont {Flohr}(2003)}]{Flohr2003}%
  \BibitemOpen
  \bibfield  {author} {\bibinfo {author} {\bibfnamefont {M.~A.~I.}\
  \bibnamefont {Flohr}},\ }\href {\doibase 10.1142/S0217751X03016859}
  {\bibfield  {journal} {\bibinfo  {journal} {International Journal of Modern
  Physics A}\ }\textbf {\bibinfo {volume} {18}},\ \bibinfo {pages} {4497}
  (\bibinfo {year} {2003})},\ \Eprint {http://arxiv.org/abs/hep-th/0111228}
  {hep-th/0111228} \BibitemShut {NoStop}%
\bibitem [{\citenamefont {Creutzig}\ and\ \citenamefont
  {Ridout}(2013)}]{CreutzigRidout2013}%
  \BibitemOpen
  \bibfield  {author} {\bibinfo {author} {\bibfnamefont {T.}~\bibnamefont
  {Creutzig}}\ and\ \bibinfo {author} {\bibfnamefont {D.}~\bibnamefont
  {Ridout}},\ }\href {\doibase 10.1088/1751-8113/46/49/494006} {\bibfield
  {journal} {\bibinfo  {journal} {Journal of Physics A: Mathematical and
  Theoretical}\ }\textbf {\bibinfo {volume} {46}},\ \bibinfo {pages} {494006}
  (\bibinfo {year} {2013})},\ \Eprint {http://arxiv.org/abs/1303.0847}
  {arXiv:1303.0847 [hep-th]} \BibitemShut {NoStop}%
\bibitem [{\citenamefont {Bender}\ \emph {et~al.}(2005)\citenamefont {Bender},
  \citenamefont {Jones},\ and\ \citenamefont {Rivers}}]{BenderJonesRivers2005}%
  \BibitemOpen
  \bibfield  {author} {\bibinfo {author} {\bibfnamefont {C.~M.}\ \bibnamefont
  {Bender}}, \bibinfo {author} {\bibfnamefont {H.~F.}\ \bibnamefont {Jones}}, \
  and\ \bibinfo {author} {\bibfnamefont {R.~J.}\ \bibnamefont {Rivers}},\
  }\href {\doibase 10.1016/j.physletb.2005.08.087} {\bibfield  {journal}
  {\bibinfo  {journal} {Phys. Lett. B}\ }\textbf {\bibinfo {volume} {625}},\
  \bibinfo {pages} {333} (\bibinfo {year} {2005})},\ \Eprint
  {http://arxiv.org/abs/hep-th/0508105} {arXiv:hep-th/0508105 [hep-th]}
  \BibitemShut {NoStop}%
\bibitem [{\citenamefont {Alexandre}\ \emph {et~al.}(2017)\citenamefont
  {Alexandre}, \citenamefont {Millington},\ and\ \citenamefont
  {Seynaeve}}]{Alexandre:2017foi}%
  \BibitemOpen
  \bibfield  {author} {\bibinfo {author} {\bibfnamefont {J.}~\bibnamefont
  {Alexandre}}, \bibinfo {author} {\bibfnamefont {P.}~\bibnamefont
  {Millington}}, \ and\ \bibinfo {author} {\bibfnamefont {D.}~\bibnamefont
  {Seynaeve}},\ }\href {\doibase 10.1103/PhysRevD.96.065027} {\bibfield
  {journal} {\bibinfo  {journal} {Phys. Rev. D}\ }\textbf {\bibinfo {volume}
  {96}},\ \bibinfo {pages} {065027} (\bibinfo {year} {2017})},\ \Eprint
  {http://arxiv.org/abs/1707.01057} {arXiv:1707.01057 [hep-th]} \BibitemShut
  {NoStop}%
\bibitem [{\citenamefont {D{\'o}ra}\ \emph {et~al.}(2022)\citenamefont
  {D{\'o}ra}, \citenamefont {Sticlet},\ and\ \citenamefont
  {Moca}}]{Dora:2021cxu}%
  \BibitemOpen
  \bibfield  {author} {\bibinfo {author} {\bibfnamefont {B.}~\bibnamefont
  {D{\'o}ra}}, \bibinfo {author} {\bibfnamefont {D.}~\bibnamefont {Sticlet}}, \
  and\ \bibinfo {author} {\bibfnamefont {C.~P.}\ \bibnamefont {Moca}},\ }\href
  {\doibase 10.1103/PhysRevLett.128.146804} {\bibfield  {journal} {\bibinfo
  {journal} {Phys. Rev. Lett.}\ }\textbf {\bibinfo {volume} {128}},\ \bibinfo
  {pages} {146804} (\bibinfo {year} {2022})},\ \Eprint
  {http://arxiv.org/abs/2112.08294} {arXiv:2112.08294 [cond-mat.str-el]}
  \BibitemShut {NoStop}%
\bibitem [{SM()}]{SM}%
  \BibitemOpen
  \href@noop {} {\bibinfo  {journal} {Supplemental Material}\ }\BibitemShut
  {NoStop}%
\bibitem [{\citenamefont {Benoit}\ and\ \citenamefont
  {Saint-Aubin}(1988)}]{BenoitSaintAubin1988}%
  \BibitemOpen
\bibfield  {journal} {  }\bibfield  {author} {\bibinfo {author} {\bibfnamefont
  {L.}~\bibnamefont {Benoit}}\ and\ \bibinfo {author} {\bibfnamefont
  {Y.}~\bibnamefont {Saint-Aubin}},\ }\href {\doibase
  10.1016/0370-2693(88)91352-4} {\bibfield  {journal} {\bibinfo  {journal}
  {Physics Letters B}\ }\textbf {\bibinfo {volume} {215}},\ \bibinfo {pages}
  {517} (\bibinfo {year} {1988})}\BibitemShut {NoStop}%
\bibitem [{\citenamefont {Kausch}()}]{Kausch1995}%
  \BibitemOpen
  \bibfield  {author} {\bibinfo {author} {\bibfnamefont {H.~G.}\ \bibnamefont
  {Kausch}},\ }\href@noop {} {\ }\Eprint {http://arxiv.org/abs/hep-th/9510149}
  {arXiv:hep-th/9510149 [hep-th]} \BibitemShut {NoStop}%
\bibitem [{\citenamefont {Kausch}(2000)}]{Kausch2000}%
  \BibitemOpen
  \bibfield  {author} {\bibinfo {author} {\bibfnamefont {H.~G.}\ \bibnamefont
  {Kausch}},\ }\href {\doibase 10.1016/S0550-3213(00)00295-9} {\bibfield
  {journal} {\bibinfo  {journal} {Nuclear Physics B}\ }\textbf {\bibinfo
  {volume} {583}},\ \bibinfo {pages} {513} (\bibinfo {year} {2000})},\ \Eprint
  {http://arxiv.org/abs/hep-th/0003029} {arXiv:hep-th/0003029} \BibitemShut
  {NoStop}%
\bibitem [{\citenamefont {Ryu}\ and\ \citenamefont {Yoon}(2023)}]{Ryu:2022ffg}%
  \BibitemOpen
  \bibfield  {author} {\bibinfo {author} {\bibfnamefont {S.}~\bibnamefont
  {Ryu}}\ and\ \bibinfo {author} {\bibfnamefont {J.}~\bibnamefont {Yoon}},\
  }\href {\doibase 10.1103/PhysRevLett.130.241602} {\bibfield  {journal}
  {\bibinfo  {journal} {Phys. Rev. Lett.}\ }\textbf {\bibinfo {volume} {130}},\
  \bibinfo {pages} {241602} (\bibinfo {year} {2023})},\ \Eprint
  {http://arxiv.org/abs/2208.12169} {arXiv:2208.12169 [hep-th]} \BibitemShut
  {NoStop}%
\bibitem [{\citenamefont {Gaberdiel}\ and\ \citenamefont
  {Kausch}(1996)}]{GaberdielKausch1996}%
  \BibitemOpen
  \bibfield  {author} {\bibinfo {author} {\bibfnamefont {M.~R.}\ \bibnamefont
  {Gaberdiel}}\ and\ \bibinfo {author} {\bibfnamefont {H.~G.}\ \bibnamefont
  {Kausch}},\ }\href {\doibase 10.1016/0550-3213(96)00364-1} {\bibfield
  {journal} {\bibinfo  {journal} {Nuclear Physics B}\ }\textbf {\bibinfo
  {volume} {477}},\ \bibinfo {pages} {293} (\bibinfo {year} {1996})},\ \Eprint
  {http://arxiv.org/abs/hep-th/9604026} {arXiv:hep-th/9604026 [hep-th]}
  \BibitemShut {NoStop}%
\bibitem [{\citenamefont {Vasseur}\ \emph {et~al.}(2011)\citenamefont
  {Vasseur}, \citenamefont {Jacobsen},\ and\ \citenamefont
  {Saleur}}]{VasseurJacobsenSaleur2011}%
  \BibitemOpen
  \bibfield  {author} {\bibinfo {author} {\bibfnamefont {R.}~\bibnamefont
  {Vasseur}}, \bibinfo {author} {\bibfnamefont {J.~L.}\ \bibnamefont
  {Jacobsen}}, \ and\ \bibinfo {author} {\bibfnamefont {H.}~\bibnamefont
  {Saleur}},\ }\href {\doibase 10.1016/j.nuclphysb.2011.05.018} {\bibfield
  {journal} {\bibinfo  {journal} {Nuclear Physics B}\ }\textbf {\bibinfo
  {volume} {851}},\ \bibinfo {pages} {314} (\bibinfo {year} {2011})},\ \Eprint
  {http://arxiv.org/abs/1103.3134} {1103.3134} \BibitemShut {NoStop}%
\bibitem [{\citenamefont {Vasseur}(2013)}]{VasseurThesis2013}%
  \BibitemOpen
  \bibfield  {author} {\bibinfo {author} {\bibfnamefont {R.}~\bibnamefont
  {Vasseur}},\ }\href {https://theses.fr/2013PA066444} {\bibinfo {type} {Phd
  thesis}},\ \bibinfo  {school} {Universit{\'e} Pierre et Marie Curie (Paris
  VI)} (\bibinfo {year} {2013})\BibitemShut {NoStop}%
\bibitem [{\citenamefont {Ashida}\ and\ \citenamefont
  {Ueda}(2018)}]{AshidaUeda2018}%
  \BibitemOpen
  \bibfield  {author} {\bibinfo {author} {\bibfnamefont {Y.}~\bibnamefont
  {Ashida}}\ and\ \bibinfo {author} {\bibfnamefont {M.}~\bibnamefont {Ueda}},\
  }\href {\doibase 10.1103/PhysRevLett.120.185301} {\bibfield  {journal}
  {\bibinfo  {journal} {Phys. Rev. Lett.}\ }\textbf {\bibinfo {volume} {120}},\
  \bibinfo {pages} {185301} (\bibinfo {year} {2018})},\ \Eprint
  {http://arxiv.org/abs/1709.03704} {arXiv:1709.03704 [cond-mat.quant-gas]}
  \BibitemShut {NoStop}%
\bibitem [{\citenamefont {Meden}\ \emph {et~al.}(2023)\citenamefont {Meden},
  \citenamefont {Grunwald},\ and\ \citenamefont
  {Kennes}}]{MedenGrunwaldKennes2023}%
  \BibitemOpen
  \bibfield  {author} {\bibinfo {author} {\bibfnamefont {V.}~\bibnamefont
  {Meden}}, \bibinfo {author} {\bibfnamefont {L.}~\bibnamefont {Grunwald}}, \
  and\ \bibinfo {author} {\bibfnamefont {D.~M.}\ \bibnamefont {Kennes}},\
  }\href {\doibase 10.1088/1361-6633/ad05f3} {\bibfield  {journal} {\bibinfo
  {journal} {Rep. Prog. Phys.}\ }\textbf {\bibinfo {volume} {86}},\ \bibinfo
  {pages} {124501} (\bibinfo {year} {2023})},\ \Eprint
  {http://arxiv.org/abs/2303.05956} {arXiv:2303.05956 [quant-ph]} \BibitemShut
  {NoStop}%
\bibitem [{\citenamefont {Lieu}(2018)}]{Lieu2018}%
  \BibitemOpen
  \bibfield  {author} {\bibinfo {author} {\bibfnamefont {S.}~\bibnamefont
  {Lieu}},\ }\href {\doibase 10.1103/PhysRevB.97.045106} {\bibfield  {journal}
  {\bibinfo  {journal} {Phys. Rev. B}\ }\textbf {\bibinfo {volume} {97}},\
  \bibinfo {pages} {045106} (\bibinfo {year} {2018})},\ \Eprint
  {http://arxiv.org/abs/1709.03788} {arXiv:1709.03788 [cond-mat.mes-hall]}
  \BibitemShut {NoStop}%
\bibitem [{\citenamefont {Koo}\ and\ \citenamefont
  {Saleur}(1994)}]{KooSaleur1994}%
  \BibitemOpen
  \bibfield  {author} {\bibinfo {author} {\bibfnamefont {W.~M.}\ \bibnamefont
  {Koo}}\ and\ \bibinfo {author} {\bibfnamefont {H.}~\bibnamefont {Saleur}},\
  }\href {\doibase 10.1016/0550-3213(94)90018-3} {\bibfield  {journal}
  {\bibinfo  {journal} {Nucl. Phys. B}\ }\textbf {\bibinfo {volume} {426}},\
  \bibinfo {pages} {459} (\bibinfo {year} {1994})},\ \Eprint
  {http://arxiv.org/abs/hep-th/9312156} {arXiv:hep-th/9312156} \BibitemShut
  {NoStop}%
\bibitem [{\citenamefont {Milsted}\ and\ \citenamefont
  {Vidal}(2017)}]{MilstedVidal2017}%
  \BibitemOpen
  \bibfield  {author} {\bibinfo {author} {\bibfnamefont {A.}~\bibnamefont
  {Milsted}}\ and\ \bibinfo {author} {\bibfnamefont {G.}~\bibnamefont
  {Vidal}},\ }\href {\doibase 10.1103/PhysRevB.96.245105} {\bibfield  {journal}
  {\bibinfo  {journal} {Phys. Rev. B}\ }\textbf {\bibinfo {volume} {96}},\
  \bibinfo {pages} {245105} (\bibinfo {year} {2017})},\ \Eprint
  {http://arxiv.org/abs/1706.01436} {arXiv:1706.01436 [cond-mat.str-el]}
  \BibitemShut {NoStop}%
\bibitem [{\citenamefont {Di~Francesco}\ \emph {et~al.}(1997)\citenamefont
  {Di~Francesco}, \citenamefont {Mathieu},\ and\ \citenamefont
  {S{\'e}n{\'e}chal}}]{DiFrancescoMathieuSenechal1997}%
  \BibitemOpen
  \bibfield  {author} {\bibinfo {author} {\bibfnamefont {P.}~\bibnamefont
  {Di~Francesco}}, \bibinfo {author} {\bibfnamefont {P.}~\bibnamefont
  {Mathieu}}, \ and\ \bibinfo {author} {\bibfnamefont {D.}~\bibnamefont
  {S{\'e}n{\'e}chal}},\ }\href {\doibase 10.1007/978-1-4612-2256-9} {\emph
  {\bibinfo {title} {Conformal Field Theory}}},\ Graduate Texts in Contemporary
  Physics\ (\bibinfo  {publisher} {Springer},\ \bibinfo {address} {New York,
  NY},\ \bibinfo {year} {1997})\BibitemShut {NoStop}%
\bibitem [{\citenamefont {Belavin}\ \emph {et~al.}(1984)\citenamefont
  {Belavin}, \citenamefont {Polyakov},\ and\ \citenamefont
  {Zamolodchikov}}]{BelavinPolyakovZamolodchikov1984}%
  \BibitemOpen
  \bibfield  {author} {\bibinfo {author} {\bibfnamefont {A.~A.}\ \bibnamefont
  {Belavin}}, \bibinfo {author} {\bibfnamefont {A.~M.}\ \bibnamefont
  {Polyakov}}, \ and\ \bibinfo {author} {\bibfnamefont {A.~B.}\ \bibnamefont
  {Zamolodchikov}},\ }\href {\doibase 10.1016/0550-3213(84)90052-X} {\bibfield
  {journal} {\bibinfo  {journal} {Nucl. Phys. B}\ }\textbf {\bibinfo {volume}
  {241}},\ \bibinfo {pages} {333} (\bibinfo {year} {1984})}\BibitemShut
  {NoStop}%
\bibitem [{\citenamefont {Polchinski}(1988)}]{Polchinski1988}%
  \BibitemOpen
  \bibfield  {author} {\bibinfo {author} {\bibfnamefont {J.}~\bibnamefont
  {Polchinski}},\ }\href {\doibase 10.1016/0550-3213(88)90179-4} {\bibfield
  {journal} {\bibinfo  {journal} {Nucl. Phys. B}\ }\textbf {\bibinfo {volume}
  {303}},\ \bibinfo {pages} {226} (\bibinfo {year} {1988})}\BibitemShut
  {NoStop}%
\bibitem [{\citenamefont {Nakayama}(2015)}]{Nakayama2015}%
  \BibitemOpen
  \bibfield  {author} {\bibinfo {author} {\bibfnamefont {Y.}~\bibnamefont
  {Nakayama}},\ }\href {\doibase 10.1016/j.physrep.2014.12.003} {\bibfield
  {journal} {\bibinfo  {journal} {Phys. Rept.}\ }\textbf {\bibinfo {volume}
  {569}},\ \bibinfo {pages} {1} (\bibinfo {year} {2015})},\ \Eprint
  {http://arxiv.org/abs/1302.0884} {arXiv:1302.0884 [hep-th]} \BibitemShut
  {NoStop}%
\end{thebibliography}%

\clearpage
\onecolumngrid

\begin{center}
{\large\bfseries Supplemental Material}
\end{center}

\begingroup
\SMredirecttoc
\setcounter{tocdepth}{2} 
\SMtableofcontents

\setcounter{secnumdepth}{1}

\setcounter{section}{0}
\renewcommand{\thesection}{\Roman{section}}

\setcounter{equation}{0}
\renewcommand{\theequation}{S\arabic{equation}}
\setcounter{table}{0}
\renewcommand{\thetable}{S\arabic{table}}
\setcounter{figure}{0}
\renewcommand{\thefigure}{S\arabic{figure}}

\section{Canonical quantization of 1+1d PT-symmetric free-fermion field theory}
\label{Canonical quantization of 1+1d PT-symmetric free-fermion field theory}

We start from the PT-symmetric non-Hermitian free-fermion field theory
introduced in the main text:
\begin{align}
S = \int_{-\infty}^{\infty} dt\int_0^{L} dx\,
\left[i\psi_+^\dagger\left(\partial_t+v\partial_x\right)\psi_+
+i\psi_-^\dagger\left(\partial_t-v\partial_x\right)\psi_-
-\Delta\psi_+^\dagger\psi_-
-u(\partial_x\psi_-^{\dagger})(\partial_x\psi_+)\right].
\end{align}
Since the action is non-Hermitian, the Euler-Lagrange equations obtained from the
functional derivatives with respect to the Grassmann fields $\psi_{\pm}^{\dagger}$ and $\psi_{\pm}$ are inequivalent. Denoting the corresponding
solutions by $\psi_\pm^{L}$ and $(\psi_\pm^{R})^{\dagger}$ and $\delta_{L/R}$ by left/right derivatives,
we obtain
\begin{align}
\label{eq:SM_EOM}
\frac{\delta_L S}{\delta\psi^{\dagger}_{\pm}}=0\ \Longrightarrow\
\left\{
\begin{aligned}
 i(\partial_t+v\partial_x)\psi^L_+-\Delta\psi^L_-&=0,\\
 i(\partial_t-v\partial_x)\psi^L_-+u\partial_x^2\psi^L_+&=0,
\end{aligned}
\right.
\qquad
\frac{\delta_R S}{\delta\psi_{\pm}}=0\ \Longrightarrow\
\left\{
\begin{aligned}
 -i(\partial_t+v\partial_x)(\psi_+^R)^\dagger +u\partial_x^2(\psi_-^R)^\dagger&=0,\\
-i(\partial_t-v\partial_x)(\psi_-^R)^\dagger-\Delta(\psi_+^R)^\dagger&=0.
\end{aligned}
\right.
\end{align}
With periodic boundary conditions on a circle of circumference $L$, the momenta are
$p_m=2\pi m/L$ with $m\in\mathbb{Z}$. The dispersion relation following from
\eqref{eq:SM_EOM} is $\omega^2=v_F^2 p^2$ with
\begin{align}
v_F=\sqrt{v^2+u\Delta},
\end{align}
where the sign choice can be absorbed into the mode conventions.
A convenient mode expansion solving \eqref{eq:SM_EOM} is
\begin{align}
\label{eq:SM_mode_expansions}
\psi_+^L(x,t)=&\frac{1}{\sqrt{L}}\Big(\psi^L_{+,0}-i\Delta t\,\psi_{-,0}^L\Big)
+\frac{1}{\sqrt{L}}\sum_{m\neq0}\Big[
B_m^L e^{ip_m(x-v_F t)}+C_m^L e^{ip_m(x+v_F t)}
\Big],
\nonumber\\
\psi_-^L(x,t)=&\frac{1}{\sqrt{L}}\psi^L_{-,0}
+\frac{1}{\sqrt{L}}\sum_{m\neq0}\frac{p_m}{\Delta}
\Big[
(v_F-v)B_m^L e^{ip_m(x-v_F t)}-(v_F+v)C_m^L e^{ip_m(x+v_F t)}
\Big],
\nonumber\\
(\psi_+^R)^\dagger(x,t)=&\frac{1}{\sqrt{L}}(\psi^R_{+,0})^\dagger
+\frac{1}{2v_F\sqrt{L}}\sum_{m\neq0}
\Big[
(v_F+v)(B_m^R)^\dagger e^{-ip_m(x-v_F t)}+(v_F-v)(C_m^R)^\dagger e^{-ip_m(x+v_F t)}
\Big],
\nonumber\\
(\psi_-^R)^\dagger(x,t)=&\frac{1}{\sqrt{L}}\Big[(\psi^R_{-,0})^\dagger+i\Delta t\,(\psi^R_{+,0})^\dagger\Big]
+\frac{\Delta}{2v_F\sqrt{L}}\sum_{m\neq0}\frac{1}{p_m}
\Big[
(B_m^R)^\dagger e^{-ip_m(x-v_F t)}-(C_m^R)^\dagger e^{-ip_m(x+v_F t)}
\Big].
\end{align}

Canonical quantization is performed in the biorthogonal sense:
\begin{align}
\label{eq:SM_ETAC}
\left\{\psi_\alpha^L(x,t),[\psi_\beta^R(y,t)]^\dagger\right\}
=\delta_{\alpha\beta}\,\delta(x-y), 
\qquad
&\left\{\psi^L_\alpha(x,t),\psi_\beta^R(y,t)\right\}=
\left\{[\psi^{L}_{\alpha}(x,t)]^{\dagger},[\psi^{R}_{\beta}(y,t)]^{\dagger}\right\}=0.
\end{align}
The mode operators $B_m^{(L/R)}$, $C_m^{(L/R)}$, and $\psi_{\pm,0}^{(L/R)}$ then obey the anti-commutation relations:
\begin{align}
\left\{B_m^L,(B_n^R)^\dagger\right\}
=\left\{C_m^L,(C_n^R)^\dagger\right\}
=\delta_{mn},
\qquad
\left\{\psi^L_{\alpha,0},(\psi_{\beta,0}^R)^\dagger\right\}
=\delta_{\alpha\beta},
\end{align}
with all others vanishing.
In terms of these modes, the Hamiltonian and linear momentum take the forms
\begin{align}
\label{eq:SM_H_modes}
H&=\int_0^{L} dx\left[-iv(\psi_+^R)^\dagger\partial_x\psi_+^L+iv(\psi_-^R)^\dagger\partial_x\psi_-^L
+\Delta(\psi_+^R)^\dagger\psi_-^L+u\partial_x(\psi_-^R)^\dagger\partial_x\psi_+^L\right]
\nonumber\\
&=\Delta(\psi_{+,0}^R)^\dagger\psi_{-,0}^L
+\sum_{m\neq0}v_F p_m\left[(B_m^R)^\dagger B_m^L-(C_m^R)^\dagger C_m^L\right],
\nonumber\\
P&=\int_0^{L} dx\left[-i(\psi_+^R)^\dagger\partial_x\psi_+^L-i(\psi_-^R)^\dagger\partial_x\psi_-^L\right]
=\sum_{m\neq0}p_m\left[(B_m^R)^{\dagger}B_m^L+(C_m^R)^{\dagger}C_m^L\right].
\end{align}
When $\Delta\neq0$, the Hamiltonian is not fully diagonalizable; the zero-mode term
$\Delta(\psi_{+,0}^R)^\dagger\psi_{-,0}^L$ implies a Jordan-cell structure in the ground-state sector.

Notice that the mode expansion \eqref{eq:SM_mode_expansions} makes it explicit that $\Delta=0$ is a
singular point for this choice of biorthogonal basis. Although the mode expansion is not unique, as one may perform similarity (canonical) transformations among the mode operators that preserve both the
biorthogonal equations of motion and the canonical anti-commutators, the $\Delta=0$ singularity
cannot be removed by any transformation that remains finite and invertible at $\Delta=0$.
Equivalently, any change of basis that eliminates the explicit $1/\Delta$ factors necessarily
requires a transformation with coefficients that diverge as
$\Delta\to 0$. This reflects the fact that $\Delta\neq 0$ and $\Delta=0$ correspond to genuinely
different structures of the biorthogonal mode decomposition (and, in particular, different
Jordan structures in the zero-mode sector), which are invariant under nonsingular similarity
transformations.

Because the theory is non-Hermitian, it is natural to distinguish right and left vacua.
We denote the right vacuum by $|0\rangle$ and its biorthogonal dual by $\langle 0|$ (i.e.
$\langle 0|0\rangle=1$), and we suppress explicit $R/L$ labels on the vacua. Correlation functions
throughout are understood as biorthogonal expectation values $\langle 0|(\cdots)|0\rangle
\equiv {}^{L}\!\langle 0|(\cdots)|0\rangle^{R}$.
Let $|0\rangle$ be annihilated by $\psi_{\pm,0}^L$, $B^{L}_{m>0}$, $(B^{R}_{m<0})^{\dagger}$,
$C^{L}_{m<0}$, and $(C^{R}_{m>0})^{\dagger}$. The ground-state subspace is then spanned by
\begin{align}
|0\rangle,\qquad (\psi_{+,0}^R)^\dagger|0\rangle,\qquad (\psi_{-,0}^R)^\dagger|0\rangle,\qquad
(\psi_{+,0}^R)^\dagger(\psi_{-,0}^R)^\dagger|0\rangle,
\end{align}
where $(\psi_{+,0}^R)^\dagger|0\rangle$ and $(\psi_{-,0}^R)^\dagger|0\rangle$ form a Jordan pair, since
\begin{align}
H(\psi_{+,0}^R)^\dagger|0\rangle=0,
\qquad
H(\psi_{-,0}^R)^\dagger|0\rangle=\Delta(\psi_{+,0}^R)^\dagger|0\rangle.
\end{align}
There are four types of excited single-particle states:
\begin{align}
&\text{right-moving particles created by}~(B_{m>0}^R)^{\dagger},
\nonumber\\
&\text{right-moving antiparticles created by}~B_{m<0}^L,
\nonumber\\
&\text{left-moving particles created by}~(C_{m<0}^R)^{\dagger},
\nonumber\\
&\text{left-moving antiparticles created by}~C_{m>0}^L.
\nonumber
\end{align}

\medskip

The biorthogonal two-point functions
\(
G^{RL}_{\alpha\beta}(x,t; y,t')=\langle 0|[\psi^R_\alpha(x,t)]^\dagger \psi^L_\beta(y,t')|0\rangle
\)
can be evaluated using the mode expansions:
\begin{align}
G^{RL}_{++}(x,t; y,t')
&=\frac{1}{L}+\frac{1}{2L}\sum_{m>0}\left[
\frac{v_F-v}{v_F}e^{-ip_m\left(x-y+v_F(t-t')\right)}
+\frac{v_F+v}{v_F}e^{ip_m\left(x-y-v_F(t-t')\right)}
\right]
\nonumber\\
&=\frac{1}{2L}\left\{
\frac{v_F-v}{v_F}\frac{1}{1-e^{-i\frac{2\pi}{L}\left(x-y+v_F t-v_F t'\right)}}
+\frac{v_F+v}{v_F}\frac{1}{1-e^{i\frac{2\pi}{L}\left(x-y-v_F t+v_F t'\right)}}
\right\}
\nonumber\\
&=-\frac{1}{2\pi i v_F}\frac{v_F^2(t-t')+v(x-y)}{(x-y)^2-v_F^2(t-t')^2}
+\frac{1}{2L}
+\frac{i\pi}{6v_F L^2}\left[v_F^2(t-t')-v(x-y)\right]
+\mathcal{O}(L^{-3}),
\end{align}
\begin{align}
G^{RL}_{--}(x,t; y,t')
&=\frac{1}{L}+\frac{1}{2L}\sum_{m>0}\left[
\frac{v_F+v}{v_F}e^{-ip_m\left(x-y+v_F(t-t')\right)}
+\frac{v_F-v}{v_F}e^{ip_m\left(x-y-v_F(t-t')\right)}
\right]
\nonumber\\
&=\frac{1}{2L}\left\{
\frac{v_F+v}{v_F}\frac{1}{1-e^{-i\frac{2\pi}{L}\left(x-y+v_F t-v_F t'\right)}}
+\frac{v_F-v}{v_F}\frac{1}{1-e^{i\frac{2\pi}{L}\left(x-y-v_F t+v_F t'\right)}}
\right\}
\nonumber\\
&=-\frac{1}{2\pi i v_F}\frac{v_F^2(t-t')-v(x-y)}{(x-y)^2-v_F^2(t-t')^2}
+\frac{1}{2L}
+\frac{i\pi}{6v_F L^2}\left[v_F^2(t-t')+v(x-y)\right]
+\mathcal{O}(L^{-3}),
\end{align}
\begin{align}
G^{RL}_{+-}(x,t; y,t')
&=-\frac{\pi u}{L^2v_F}\sum_{m>0}m\left\{
e^{-ip_m\left(x-y+v_F(t-t')\right)}+e^{ip_m\left(x-y-v_F(t-t')\right)}
\right\}
\nonumber\\
&=-\frac{\pi u}{L^2v_F}\left\{
\frac{e^{-i\frac{2\pi}{L}\left(x-y+v_F t-v_F t'\right)}}
{\left[1-e^{-i\frac{2\pi}{L}\left(x-y+v_F t-v_F t'\right)}\right]^2}
+\frac{e^{i\frac{2\pi}{L}\left(x-y-v_F t+v_F t'\right)}}
{\left[1-e^{i\frac{2\pi}{L}\left(x-y-v_F t+v_F t'\right)}\right]^2}
\right\}
\nonumber\\
&=
\frac{u}{4\pi v_F}\frac{1}{\left(x-y+v_F t-v_F t'\right)^2}
+\frac{u}{4\pi v_F}\frac{1}{\left(x-y-v_F t+v_F t'\right)^2}
+\frac{\pi u}{6L^2v_F}
+\mathcal{O}(L^{-3}),
\end{align}
\begin{align}
G^{RL}_{-+}(x,t; y,t')
&=\frac{i\Delta}{L}(t-t')
-\frac{\Delta}{4\pi v_F}\sum_{m>0}\frac{1}{m}\left\{
e^{-ip_m\left(x-y+v_F(t-t')\right)}+e^{ip_m\left(x-y-v_F(t-t')\right)}
\right\}
\nonumber\\
&=\frac{i\Delta}{L}(t-t')
+\frac{\Delta}{4\pi v_F}\left\{
\ln\left[1-e^{-i\frac{2\pi}{L}\left(x-y+v_F t-v_F t'\right)}\right]
+\ln\left[1-e^{i\frac{2\pi}{L}\left(x-y-v_F t+v_F t'\right)}\right]
\right\}
\nonumber\\
&=
\frac{\Delta}{4\pi v_F}\ln\left\{\frac{4\pi^2}{L^2}\left[(x-y)^2-v_F^2(t-t')^2\right]\right\}
+\frac{i\Delta}{2L}(t-t')
-\frac{\pi\Delta}{12v_F L^2}\left[v_F^2(t-t')^2+(x-y)^2\right]
+\mathcal{O}(L^{-4}).
\end{align}
We see that $G^{RL}_{-+}(x,t; y,t')$ exhibits logarithmic behavior, while the others show
power-law decay in the large-$L$ limit. In particular, taking the equal-time limit $t=t'$ gives
\begin{align}
\label{eq:SM_scaling_forms_equal_time}
G^{RL}_{++}(x,t; y,t) &\sim -\,\frac{v}{2\pi i v_F}\,\frac{1}{x-y},
\nonumber\\
G^{RL}_{--}(x,t; y,t) &\sim \phantom{-}\,\frac{v}{2\pi i v_F}\,\frac{1}{x-y},
\nonumber\\
G^{RL}_{+-}(x,t; y,t) &\sim \frac{u}{2\pi v_F}\,\frac{1}{(x-y)^2},
\nonumber\\
G^{RL}_{-+}(x,t; y,t) &\sim \frac{\Delta}{2\pi v_F}\,\ln\!\left(\frac{|x-y|}{L}\right).
\end{align}
When $\Delta=u=0$, these correlators reduce to those of the 1+1d massless Dirac fermion theory.

\section{Conformal invariance of the PT-symmetric free-fermion field theory}

\subsection{Improved energy-momentum tensor}
\label{Improved energy-momentum tensor}

To test conformal invariance, it is convenient to start from the canonical energy-momentum
tensor. For a non-Hermitian theory of Grassmann fields, we treat $\psi_\alpha$ and $\psi_\alpha^\dagger$
as independent fields and use right/left functional derivatives. The canonical energy-momentum tensor  is
\begin{align}
T_c^{\mu\nu}
=\sum_{\alpha=\pm}\left(
\frac{\partial_R\mathcal{L}}{\partial(\partial_\mu\psi_\alpha)}\,\partial^\nu\psi_\alpha
+\partial^\nu\psi_\alpha^\dagger\,\frac{\partial_L\mathcal{L}}{\partial(\partial_\mu\psi_\alpha^\dagger)}
\right)-\eta^{\mu\nu}\mathcal{L},
\end{align}
where the Minkowski metric is $\eta^{\mu\nu}=\mathrm{diag}(+1,-1)$.
For the 1+1d PT-symmetric free-fermion Lagrangian density
\begin{align}
\mathcal{L}=i\psi_+^\dagger(\partial_t+v\partial_x)\psi_+
+i\psi_-^\dagger(\partial_t-v\partial_x)\psi_-
-\Delta\,\psi_+^\dagger\psi_-
-u(\partial_x\psi_-^{\dagger})(\partial_x\psi_+),
\end{align}
the canonical energy-momentum tensor  components are
\begin{align}
T^{00}_c&=-iv\left(\psi_+^\dagger\frac{\partial\psi_+}{\partial x}-\psi_-^\dagger\frac{\partial\psi_-}{\partial x}\right)
+\Delta\psi_+^\dagger\psi_-+u\frac{\partial\psi_-^\dagger}{\partial x}\frac{\partial\psi_+}{\partial x},
\nonumber\\
T^{01}_c&=-i\psi_+^\dagger\frac{\partial\psi_+}{\partial x}-i\psi_-^\dagger\frac{\partial\psi_-}{\partial x},
\nonumber\\
T^{10}_c&=iv\left(\psi_+^\dagger\frac{\partial\psi_+}{\partial t}-\psi_-^\dagger\frac{\partial\psi_-}{\partial t}\right)
-u\frac{\partial\psi_-^\dagger}{\partial t}\frac{\partial\psi_+}{\partial x}
-u\frac{\partial\psi_-^\dagger}{\partial x}\frac{\partial\psi_+}{\partial t},
\nonumber\\
T^{11}_c&=i\left(\psi_+^\dagger\frac{\partial\psi_+}{\partial t}+\psi_-^\dagger\frac{\partial\psi_-}{\partial t}\right)
-\Delta\psi^\dagger_+\psi_-+u\frac{\partial\psi_-^\dagger}{\partial x}\frac{\partial\psi_+}{\partial x}.
\end{align}

A standard diagnostic for conformal invariance is whether one can choose a conserved energy-momentum tensor  whose
trace vanishes on-shell. One may add to the canonical energy-momentum tensor  an improvement term of the form
\begin{align}
T^{\mu\nu}=T^{\mu\nu}_c+\partial_\alpha\!\left(\varepsilon^{\mu\alpha}I^\nu\right),
\qquad
\partial_\mu T^{\mu\nu}
=\partial_\mu T^{\mu\nu}_c+\partial_\mu\partial_\alpha\!\left(\varepsilon^{\mu\alpha}I^\nu\right)
=\partial_\mu T^{\mu\nu}_c,
\end{align}
where $\varepsilon^{\mu\alpha}$ is the Levi-Civita antisymmetric tensor (with $\varepsilon^{01}=+1$)
and $I^\nu$ is a local current. The last equality follows because $\partial_\mu\partial_\alpha$ is
symmetric in $(\mu,\alpha)$ while $\varepsilon^{\mu\alpha}$ is antisymmetric, so
$\partial_\mu\partial_\alpha(\varepsilon^{\mu\alpha}I^\nu)=0$ identically.

The trace of the improved energy-momentum tensor  is
\begin{align}
T^\mu_{\ \mu}
=\left(T_c\right)^\mu_{\ \mu}+\partial_\alpha\!\left(\varepsilon^{\mu\alpha}I_\mu\right)
=\left(T_c\right)^\mu_{\ \mu}-\partial_t I_1+\partial_x I_0,
\end{align}
where $I_\mu=\eta_{\mu\nu}I^\nu$. For the present theory the canonical trace is
\begin{align}
\left(T_c\right)^\mu_{\ \mu}
=-i\psi^\dagger_+\left(\partial_t+v\partial_x\right)\psi_+
-i\psi^\dagger_-\left(\partial_t-v\partial_x\right)\psi_-
+2\Delta\psi^\dagger_+\psi_-,
\end{align}
which is not traceless even after imposing the equations of motion. This does not by itself rule out
conformal symmetry, since an appropriate improvement may exist.
To find such an improvement, we set
\begin{align}\label{eqt}
\partial_tI_1-\partial_xI_0
=-i\psi^\dagger_+\left(\partial_t+v\partial_x\right)\psi_+
-i\psi^\dagger_-\left(\partial_t-v\partial_x\right)\psi_-
+2\Delta\psi^\dagger_+\psi_-,
\end{align}
and use the equations of motion
\begin{align}
\Delta\psi^\dagger_+=-i\left(\partial_t-v\partial_x\right)\psi_-^\dagger,
\qquad
\Delta\psi_- = i\left(\partial_t+v\partial_x\right)\psi_+,
\end{align}
where we have dropped the superscript $R$ on $\psi_{\pm}^\dagger$ and $L$ on $\psi_{\pm}$ for simplicity.
Substituting these into \eqref{eqt} gives
\begin{align}
\partial_tI_1-\partial_xI_0=-i\left(\partial_t-v\partial_x\right)\left(\psi^\dagger_-\psi_-\right).
\end{align}
A simple local solution is therefore
\begin{align}
I_0=-iv\,\psi_-^\dagger\psi_-,
\qquad
I_1=-i\,\psi_-^\dagger\psi_-,
\end{align}
which makes the improved energy-momentum tensor  traceless on-shell. With this choice, the improved energy-momentum tensor  components become
\begin{align}
T^{00}&=T^{00}_c+\partial_xI_0
=-iv\left(\psi_+^\dagger\frac{\partial\psi_+}{\partial x}
+\frac{\partial\psi_-^\dagger}{\partial x}\psi_-\right)
+\Delta\psi_+^\dagger\psi_-+u\frac{\partial\psi_-^\dagger}{\partial x}\frac{\partial\psi_+}{\partial x},
\nonumber\\
T^{01}&=T^{01}_c-\partial_xI_1
=-i\left(\psi_+^\dagger\frac{\partial\psi_+}{\partial x}
-\frac{\partial\psi_-^\dagger}{\partial x}\psi_-\right),
\nonumber\\
T^{10}&=T^{10}_c-\partial_tI_0
=iv\left(\psi_+^\dagger\frac{\partial\psi_+}{\partial t}
+\frac{\partial\psi_-^\dagger}{\partial t}\psi_-\right)
-u\frac{\partial\psi_-^\dagger}{\partial t}\frac{\partial\psi_+}{\partial x}
-u\frac{\partial\psi_-^\dagger}{\partial x}\frac{\partial\psi_+}{\partial t},
\nonumber\\
T^{11}&=T^{11}_c+\partial_tI_1
=i\left(\psi_+^\dagger\frac{\partial\psi_+}{\partial t}
-\frac{\partial\psi_-^\dagger}{\partial t}\psi_-\right)
-\Delta\psi^\dagger_+\psi_-+u\frac{\partial\psi_-^\dagger}{\partial x}\frac{\partial\psi_+}{\partial x}.
\end{align}
Moreover, using the equations of motion one can check that the improved energy-momentum tensor  satisfies 
\begin{align}
T^{10}=v_F^2\,T^{01}, \qquad v_F=\sqrt{v^2+u\Delta}.
\end{align}
In the rescaled coordinates $x^0=v_F t$, the improved energy-momentum tensor  is symmetric in addition to being traceless, as expected in a 2d conformal field theory.

For comparison, one consider instead the more general bilinear theory (with $u=0$)
\begin{align}
\mathcal{L}=i\psi_+^\dagger\left(\partial_t+v\partial_x\right)\psi_+
+i\psi_-^\dagger\left(\partial_t-v\partial_x\right)\psi_-
-\Delta\psi_+^\dagger\psi_- -\Delta'\psi_-^\dagger\psi_+.
\end{align}
Following the same logic, a traceless improvement would require a local current $I_\mu$ such that
\begin{align}
\partial_t I_1-\partial_x I_0
=(T_c)^\mu_{\ \mu}\Big|_{\rm on\mbox{-}shell}
=\Delta\psi^\dagger_+\psi_-+\Delta'\psi^\dagger_-\psi_+.
\end{align}
Using the equations of motion, the right-hand side may be rewritten as
\begin{align}
\Delta\psi^\dagger_+\psi_-+\Delta'\psi^\dagger_-\psi_+
=
i\left[\psi^\dagger_-\left(\partial_t\psi_-\right)-\left(\partial_t\psi_-^\dagger\right)\psi_-\right]
-i v\left[\psi^\dagger_-\left(\partial_x\psi_-\right)-\left(\partial_x\psi_-^\dagger\right)\psi_-\right].
\end{align}
In contrast to the case with $\Delta'=0$, this on-shell trace cannot be written as
$\partial_t I_1-\partial_x I_0$ with a local dimension-one current built from fermion bilinears. Namely, the trace cannot be eliminated by a total-derivative improvement of the energy-momentum tensor .
Therefore the theory does not admit a traceless improved energy-momentum tensor  and is not conformal.
This conclusion is consistent with the spectrum: for $\Delta\Delta'\neq 0$ the dispersion relation is
\begin{align}
E(p_m)=\pm\sqrt{v^2p_m^2+\Delta\Delta'},
\end{align}
which is gapped and thus introduces an intrinsic mass scale, incompatible with conformal invariance.

\subsection{Virasoro algebra}
\label{Virasoro algebra}

Using the improved energy-momentum tensor  $T^{\mu\nu}$ constructed in the previous subsection,
we define the Virasoro operators by Fourier transforming the
(improved) energy and momentum densities. Concretely, we set
\begin{align}
\label{higher_HnPn_field}
L_{n}+\bar{L}_{-n}
&= \frac{L}{2\pi v_F}\int_{0}^{L} dx\,e^{-i\frac{2\pi n}{L}x}\,T^{00}(x)
+ \frac{c}{12}\delta_{n,0},
\nonumber\\
L_{n}-\bar{L}_{-n}
&= \frac{L}{2\pi}\int_{0}^{L} dx\,e^{-i\frac{2\pi n}{L}x}\,T^{01}(x),
\end{align}
where the constant shift with $c$  is the usual Casimir contribution on the spatial circle.

In terms of the fermionic mode operators that diagonalize the Hamiltonian and momentum, they can be written explicitly as:
\begin{align}
L_{n} &=
\begin{cases}
(B^{R}_{-n})^{\dagger}\psi^{L}_{-,0}
+ n\,(\psi^{R}_{+,0})^{\dagger} B^{L}_{n}
+ \displaystyle\sum_{m\neq 0,n}
  m\,(B^{R}_{m-n})^{\dagger} B^{L}_{m},
& n\neq 0,
\\[0.6em]
(\psi^{R}_{+,0})^{\dagger}\psi^{L}_{-,0}
+ \displaystyle\sum_{m\neq 0}
  m\,(B^{R}_{m})^{\dagger} B^{L}_{m}
+ \frac{c}{24},
& n = 0 ,
\end{cases}
\end{align}
and
\begin{align}
\bar{L}_{n} &=
\begin{cases}
(C^{R}_{n})^{\dagger}\psi^{L}_{-,0}
+ n\,(\psi^{R}_{+,0})^{\dagger}\,C^{L}_{-n}
- \displaystyle\sum_{m\neq 0,-n}
  m\,(C^{R}_{m+n})^{\dagger} C^{L}_{m},
& n\neq 0,
\\[0.6em]
(\psi^{R}_{+,0})^{\dagger}\psi^{L}_{-,0}
- \displaystyle\sum_{m\neq 0}
  m\,(C^{R}_{m})^{\dagger} C^{L}_{m}
+ \frac{c}{24},
& n = 0.
\end{cases}
\end{align}
The zero modes and oscillating modes satisfy the biorthogonal canonical anti-commutation relations
\begin{align}
\{\psi_{\alpha,0}^{L}, (\psi_{\beta,0}^{R})^{\dagger}\}
&= \delta_{\alpha\beta},
\qquad
\{B^{L}_{m}, (B^{R}_{n})^{\dagger}\}
= \{C^{L}_{m}, (C^{R}_{n})^{\dagger}\}
= \delta_{mn},
\qquad
\text{others} = 0.
\end{align}
Here we have rescaled the zero modes according to
$\frac{\Delta L}{4\pi v_F}\psi^{L}_{-,0}\rightarrow\psi^{L}_{-,0}$ and
$\bigl(\frac{\Delta L}{4\pi v_F}\bigr)^{-1}(\psi^{R}_{+,0})^{\dagger}
\rightarrow(\psi^{R}_{+,0})^{\dagger}$, which preserves their anti-commutation relation.
A fixed time slice $t=0$ is chosen for the above expressions of $L_{n}$ and $\bar{L}_{n}$; the corresponding Heisenberg-picture operators carry the expected Virasoro time dependence:
\begin{align}
L_{n}(t) = e^{-i \frac{2\pi n}{L} v_F t}\,L_{n}(0),
\qquad
\bar{L}_{n}(t)= e^{+i \frac{2\pi n}{L} v_F t}\,\bar{L}_{n}(0).
\end{align}

Using the above mode expressions and the canonical anti-commutators, one finds
\begin{align}
[L_{m}, L_{n}]
&= (m-n)\,L_{m+n}
   - \frac{2m^{3}+c\,m}{12}\,\delta_{m+n,0},
\nonumber\\
[\bar{L}_{m}, \bar{L}_{n}]
&= (m-n)\,\bar{L}_{m+n}
   - \frac{2m^{3}+c\,m}{12}\,\delta_{m+n,0},
\nonumber\\
[L_{m}, \bar{L}_{n}] &= 0.
\end{align}
Requiring the standard Virasoro central extension then fixes
\begin{align}
c=-2.
\end{align}

\section{General form of the indecomposability parameters}
\label{General form of the indecomposability parameters}

Here we derive a closed form for the indecomposability parameters $\beta_M$
associated with the holomorphic staggered Virasoro modules in our non-Hermitian
free-fermion theory.

The three right states in the level-$M$ staggered module are
\begin{align}
|\phi_M^R\rangle&=(B_M^R)^\dagger\cdots(B_1^R)^\dagger
(\psi_{+,0}^R)^\dagger|0\rangle,
\nonumber\\
|\psi_M^R\rangle&=(B_M^R)^\dagger\cdots(B_1^R)^\dagger
(\psi_{-,0}^R)^\dagger|0\rangle,
\nonumber\\
|\xi_M^R\rangle&=
\begin{cases}
(\psi_{+,0}^R)^\dagger(\psi_{-,0}^R)^\dagger|0\rangle,
\quad M=1,
\\[0.3em]
(B_{M-1}^R)^\dagger\cdots(B_1^R)^\dagger
(\psi_{+,0}^R)^\dagger(\psi_{-,0}^R)^\dagger|0\rangle,
\quad M>1.
\end{cases}
\end{align}
Let $A_{M,-}$ denote the level-$M$ singular-vector operator mapping $|\xi_M^R\rangle$ to
$|\phi_M^R\rangle$. We adopt the standard normalization in which the coefficient of
$L_{-1}^M$ in $A_{M,-}$ equals $1$. We then define the numbers $a_M$ and $b_M$ by
\begin{equation}
A_{M,-}\left|\xi_M^R\right\rangle = a_M \left|\phi_M^R\right\rangle,
\qquad
A_{M,+}\left|\psi_M^R\right\rangle = b_M \left|\xi_M^R\right\rangle,
\label{eq:def-aM-bM}
\end{equation}
and the indecomposability parameter by
\begin{equation}
\beta_M = a_M b_M.
\label{eq:def-betaM}
\end{equation}
With this definition, $\beta_M$ is invariant under rescaling of $|\xi_M^R\rangle$.

A convenient closed expression for $A_{M,-}$ is the Benoit-(Saint-Aubin) sum over
compositions of $M$~\cite{BenoitSaintAubin1988}:
\begin{equation}
A_{M,-}
= C_M^{-1}\!\!\!\sum_{\substack{\text{$(k_1,k_2,\ldots,k_r)\in$}\\ \text{compositions of $M$}}}
(-2)^{M-r}\,c(\vec{k})\,L_{-k_1}\cdots L_{-k_r},
\label{eq:AM-BSA}
\end{equation}
where the sum runs over all compositions (ordered partitions) of $M$, i.e.
$k_i\in\mathbb{Z}_{>0}$ and $k_1+\cdots+k_r=M$. The coefficient is
\begin{equation}
c(\vec{k})
:=\frac{1}{M}\prod_{i=1}^{r-1}\frac{1}{(k_1+\cdots+k_i)(k_{i+1}+\cdots+k_r)}
=\frac{1}{M}\prod_{i=1}^{r-1}\frac{1}{s_i\,(M-s_i)},
\qquad
s_i:=k_1+\cdots+k_i.
\label{eq:ck}
\end{equation}
The normalization constant is fixed by the composition $(1,1,\ldots,1)$:
\begin{equation}
C_M := c(\underbrace{1,1,\ldots,1}_{M\ \mathrm{entries}})
= \frac{1}{M[(M-1)!]^2},
\qquad\Longrightarrow\qquad
C_M^{-1}=M[(M-1)!]^2,
\label{eq:CN}
\end{equation}
so that the coefficient of $L_{-1}^M$ in $A_{M,-}$ is indeed $1$.
The biorthogonal adjoint $A_{M,+}$ is obtained from $A_{M,-}$ by the replacement
$L_{-n}\mapsto L_{n}$ and reversing the order of Virasoro modes; in particular,
each monomial $L_{-k_1}\cdots L_{-k_r}$ in $A_{M,-}$ corresponds to
$L_{k_r}\cdots L_{k_1}$ in $A_{M,+}$.

Using the explicit mode expression for $L_n$ in Eq.~\eqref{Virasoro_generator}, one obtains by
repeated application that for any composition $\vec{k}=(k_1,\ldots,k_r)$ of $M$,
\begin{align}
L_{-k_1}\cdots L_{-k_r}\left|\xi_M^R\right\rangle
&= (-1)^{M-r+1}\Big(\prod_{i=1}^{r-1} s_i\Big)\left|\phi_M^R\right\rangle,
\label{eq:Lminus-on-xi}
\\
L_{k_r}\cdots L_{k_1}\left|\psi_M^R\right\rangle
&= (-1)^{M-r}\,M\Big(\prod_{i=1}^{r-1} s_i\Big)\left|\xi_M^R\right\rangle.
\label{eq:Lplus-on-psi}
\end{align}
Inserting \eqref{eq:Lminus-on-xi} into \eqref{eq:AM-BSA} and using \eqref{eq:ck} yields
\begin{align}
A_{M,-}\left|\xi_M^R\right\rangle
&=
-\frac{C_M^{-1}2^{M}}{M}
\sum_{\vec{k}\,\vDash\,M}2^{-r}\prod_{i=1}^{r-1}\frac{1}{M-s_i}\;
\left|\phi_M^R\right\rangle,
\label{eq:AM-on-xi-sum}
\\
A_{M,+}\left|\psi_M^R\right\rangle
&=
\phantom{-}C_M^{-1}2^{M}
\sum_{\vec{k}\,\vDash\,M}2^{-r}\prod_{i=1}^{r-1}\frac{1}{M-s_i}\;
\left|\xi_M^R\right\rangle,
\label{eq:AMdag-on-psi-sum}
\end{align}
where $\vec{k}\vDash M$ denotes ``$\vec{k}$ is a composition of $M$''.
To evaluate the remaining sum, define
\begin{equation}
u_i := M-s_{r-i},\qquad i=1,\ldots,r-1.
\end{equation}
Then $1\le u_1<\cdots<u_{r-1}\le M-1$, and the map
$\vec{k}\mapsto S(\vec{k}):=\{u_1,\ldots,u_{r-1}\}\subset\{1,\ldots,M-1\}$
is a bijection between compositions of $M$ and subsets of $\{1,\ldots,M-1\}$
(with $r=|S|+1$). Therefore,
\begin{align}
\sum_{\vec{k}\,\vDash\,M}2^{-r}\prod_{i=1}^{r-1}\frac{1}{M-s_i}
&=
\sum_{S\subset\{1,\ldots,M-1\}}2^{-|S|-1}\prod_{u\in S}\frac{1}{u}
=
\frac12\sum_{S\subset\{1,\ldots,M-1\}}\prod_{u\in S}\frac{1}{2u}
\nonumber\\
&=
\frac12\prod_{u=1}^{M-1}\left(1+\frac{1}{2u}\right)
=
\frac{(2M-1)!}{2^{2M-1}[(M-1)!]^2}.
\label{eq:composition-sum}
\end{align}

Using \eqref{eq:CN}, \eqref{eq:AM-on-xi-sum}, \eqref{eq:AMdag-on-psi-sum}, and
\eqref{eq:composition-sum}, we obtain from \eqref{eq:def-aM-bM} the closed forms
\begin{equation}
a_M=-\frac{(2M-1)!}{2^{M-1}},
\qquad
b_M=\frac{M(2M-1)!}{2^{M-1}},
\qquad
\beta_M=a_M b_M=-\frac{M\big[(2M-1)!\big]^2}{4^{\,M-1}}.
\label{eq:betaM-closed-form}
\end{equation}

\section{LCFT data of the PT-symmetric free-fermion lattice model}
\label{LCFT data of the PT-symmetric free-fermion lattice model}

\subsection{Spectrum and two-point correlation functions at EP}
\label{subsec:spec_corr_EP}

We start from the lattice Hamiltonian $H_{\mathrm{TB}}=\sum_{j=1}^{N}h_j$ with tight-binding density
\begin{align}
h_j=-[t+(-1)^j\delta]\,(c^\dagger_{j+1}c_j+c^\dagger_j c_{j+1})-(-1)^j\mu_s\,c^\dagger_j c_j,
\label{eq:latt_hj}
\end{align}
and specialize to the EP $\mu_s=2i\delta$,
\begin{align}
H_{\mathrm{EP}}
:=H_{\mathrm{TB}}\big|_{\mu_s=2i\delta}.
\end{align}
Here we assume periodic boundary condition and even $N$.
For convenience, we also set the lattice spacing $a=1$, so that the system length is $L=N$.

\paragraph{Single-particle spectrum.}
We first take a Fourier transform of the site fermions:
\begin{align}
c_j=\frac{1}{\sqrt{N}}\sum_{p} e^{ipj}\,\theta_p,
\qquad
c_j^\dagger=\frac{1}{\sqrt{N}}\sum_{p} e^{-ipj}\,\theta^\dagger_{p},
\label{eq:FT_theta}
\end{align}
with $p\in\frac{2\pi}{N}\mathbb{Z}$.  Because the Hamiltonian has period two, modes at momenta
differing by $\pi$ are coupled.  It is therefore convenient to fold the Brillouin zone and work with
the reduced momenta
\begin{align}
p_m=\frac{2\pi m}{N},\qquad m=1,2,\dots,\frac{N}{2},
\end{align}
and to introduce the two-component spinor
$
\Theta(p_m):=\big(\theta_{p_m-\frac{\pi}{2}},\,\theta_{p_m+\frac{\pi}{2}}\big)^{\mathsf{T}}.
$
The Hamiltonian can then be written as
\begin{align}\label{SSHbloch}
H_{\mathrm{EP}}
&=\sum_{m=1}^{N/2}
\begin{pmatrix}
\theta^\dagger_{p_m-\pi/2} & \theta^\dagger_{p_m+\pi/2}
\end{pmatrix}
\mathcal{H}(p_m)
\begin{pmatrix}
\theta_{p_m-\pi/2} \\
\theta_{p_m+\pi/2}
\end{pmatrix},
\nonumber\\
\mathcal{H}(p)
&=
\begin{pmatrix}
-2t\sin p & -2i\delta(1+\cos p)  \\
-2i\delta(1-\cos p) & 2t\sin p
\end{pmatrix}.
\end{align}

For $p\neq \pi$, the Bloch Hamiltonian $\mathcal{H}(p)$ is diagonalizable.
Because $\mathcal{H}(p)$ is non-Hermitian, we diagonalize it in a biorthogonal basis.  We choose
matrices $S_R(p)$ and $S_L(p)$ such that
\begin{align}
S_L^\dagger(p)\,\mathcal{H}(p)\,S_R(p)=
\begin{pmatrix}
E(p)&0\\
0&-E(p)
\end{pmatrix},
\qquad
S_L^\dagger(p)\,S_R(p)=\mathbbm{1}.
\label{eq:biorth_diag}
\end{align}
Here $E(p)=v_F\sin p$, with $v_F=2\sqrt{t^2-\delta^2}$.
A convenient choice for the right eigenvector matrix is
\begin{align}
S_R(p)=\frac{1}{\sqrt{2}}
\begin{pmatrix}
1&1\\[0.25em]
-\dfrac{v_F+2t}{2i\delta}\tan\!\left(\dfrac{p}{2}\right)
&
+\dfrac{v_F-2t}{2i\delta}\tan\!\left(\dfrac{p}{2}\right)
\end{pmatrix},
\label{eq:UR_def}
\end{align}
and $S_L^\dagger(p)=S_R(p)^{-1}$.
We then define the left/right eigenmode operators by
\begin{align}
\begin{pmatrix}
\chi_L(p)\\ \eta_L(p)
\end{pmatrix}
&:=S_L^\dagger(p)
\begin{pmatrix}
\theta_{p-\pi/2}\\ \theta_{p+\pi/2}
\end{pmatrix},
\qquad
\begin{pmatrix}
\chi_R(p)\\ \eta_R(p)
\end{pmatrix}
:=S_R^\dagger(p)
\begin{pmatrix}
\theta_{p-\pi/2}\\ \theta_{p+\pi/2}
\end{pmatrix},
\label{eq:eigenmodes_def}
\end{align}
which satisfy the biorthogonal anti-commutation relations
\begin{align}
\{\chi_L(p_m),\chi_R^\dagger(p_n)\}
=\{\eta_L(p_m),\eta_R^\dagger(p_n)\}
=\delta_{mn},
\qquad
\text{(others vanish)}.
\label{eq:biorth_ACR_modes}
\end{align}

At $p=\pi$, the Bloch Hamiltonian becomes defective,
\begin{align}
\mathcal{H}(\pi)=
\begin{pmatrix}
0&0\\
-4i\delta&0
\end{pmatrix},
\end{align}
forming a rank-one Jordan block.  This involves the pair $\theta_{\pi/2}$ and $\theta_{3\pi/2}$, and
we keep these modes explicitly by defining
\begin{align}
\chi_0:=\theta_{3\pi/2},\qquad \eta_0:=\theta_{\pi/2},
\end{align}
so that the $p=\pi$ contribution to the Hamiltonian is
\begin{align}
H_{0}=\alpha_0\,\chi_0^\dagger \eta_0,\qquad \alpha_0=-4i\delta.
\label{zeromode_prefactor}
\end{align}

Collecting the diagonalizable blocks ($m=1,\dots,\frac{N}{2}-1$) and the Jordan block ($m=\frac{N}{2}$) , we obtain
\begin{align}
H_{\mathrm{EP}}
=\alpha_0\,\chi_0^\dagger\eta_0
+\sum_{m=1}^{N/2-1}E(p_m)\Big[
\chi_R^\dagger(p_m)\chi_L(p_m)
-\eta_R^\dagger(p_m)\eta_L(p_m)
\Big].
\label{eq:HEP_modes}
\end{align}

\paragraph{Two-point correlation functions.}
Let $|0\rangle$ be the Fock vacuum annihilated by all $\theta_p$ (equivalently, by all $\chi_R(p)$,
$\eta_R(p)$, and by $\chi_0$ and $\eta_0$). We define the right and left ground states by filling
the negative-energy modes, excluding the zero modes:
\begin{align}
|GS_R\rangle:=\prod_{m=1}^{N/2-1}\eta_R^\dagger(p_m)\,|0\rangle,
\qquad
\langle GS_L|:=\langle 0|\prod_{m=1}^{N/2-1}\eta_L(p_m),
\label{eq:GS_biorth}
\end{align}
with a fixed choice of ordering in the products (e.g.\ increasing $m$).  The equal-time lattice
two-point function in the biorthogonal formalism is then
\begin{align}
\langle c^\dagger_jc_{j'}\rangle
&:=\langle GS_L|c^\dagger_jc_{j'}|GS_R\rangle
\nonumber\\
&=\frac{1}{N}\,i^{\,j-j'}\sum_{m=1}^{N/2-1}e^{ip_m(j'-j)}
\Bigg[
\frac{v_F+2t}{2v_F}
+(-1)^{j-j'}\frac{v_F-2t}{2v_F}
+(-1)^j\frac{i\delta}{v_F}\cot\!\left(\frac{p_m}{2}\right)
+(-1)^{j'}\frac{i\delta}{v_F}\tan\!\left(\frac{p_m}{2}\right)
\Bigg].
\label{eq:corr_latt_exactsum}
\end{align}
In the thermodynamic limit, the dominant long-distance contribution comes from the singular behavior
$\cot(p/2)\sim 2/p$ as $p\to 0$, which yields a harmonic sum and hence a logarithm.  Consequently,
for $1\ll |j-j'|\ll N$ one finds the logarithmic scaling
\begin{align}
\langle c^\dagger_j c_{j'}\rangle
\;\sim\;
(-1)^j\, i^{\,j-j'}\,\kappa\,
\ln\!\left(\frac{|j-j'|}{N}\right),
\end{align}
where $\kappa$ is a nonuniversal amplitude proportional to $\delta/v_F$.
This agrees with the continuum prediction~\eqref{eq:SM_scaling_forms_equal_time}, up to the expected
oscillatory phase factors fixed by the lattice--continuum embedding.

\subsection{Lattice Virasoro operators and indecomposability parameters}
\label{subsec:latt_Vir_beta}

Now we follow the Koo-Saleur construction~\cite{KooSaleur1994} to characterize the
conformal structure of the EP lattice model in the scaling limit $L=N\to\infty$ (as we set the
lattice spacing $a=1$).  The central idea is to form appropriate Fourier modes of the local energy
and momentum densities, which converge to (linear combinations of) Virasoro generators in the
continuum.

\medskip

\paragraph{Koo-Saleur construction.}
Given a local Hamiltonian $H=\sum_{j=1}^N h_j$, the Koo-Saleur prescription defines the
corresponding momentum density (motivated by locality and energy--momentum conservation) as
\begin{align}
\label{lattice_pj}
p_j := i[h_j, h_{j+1}],
\end{align}
and introduces the Fourier modes
\begin{align}
\label{lattice_HnPn}
H^{\mathrm{latt}}_n
&=L^{\mathrm{latt}}_n+\bar L^{\mathrm{latt}}_{-n}
:=\frac{Na}{2\pi v_F}\sum_{j=1}^N e^{-ip_n j}\left(h_j-h_\infty\right)
+\frac{c}{12}\delta_{n,0},
\nonumber\\
P^{\mathrm{latt}}_n
&=L^{\mathrm{latt}}_n-\bar L^{\mathrm{latt}}_{-n}
:=\frac{Na^2}{2\pi v_F^2}\sum_{j=1}^N e^{-ip_n j}\,p_j,
\end{align}
where $p_n:=2\pi n/N$, $v_F$ is the Fermi velocity, $c$ is the central charge, and
$h_\infty=\lim_{N\to\infty}E_{\mathrm{GS}}(N)/N$ is the ground-state energy density.
For simplicity we drop the superscript ``latt'' below.  The holomorphic and antiholomorphic
generators are obtained from
\begin{align}
L_n=\frac{H_n+P_n}{2},\qquad \bar L_{-n}=\frac{H_n-P_n}{2}.
\label{eq:Lbar_from_HP}
\end{align}

For the EP lattice model with the bare density $h_j$ in Eq.~\eqref{eq:latt_hj}, a direct
implementation of the Koo-Saleur construction does not reproduce the expected Virasoro algebra in
the scaling limit.  Concretely, while $H_0$ correctly yields the Hamiltonian, the higher modes built
from $h_j$ fail to converge to the continuum Virasoro generators, e.g., the commutators
$[L_n,L_m]-(n-m)L_{n+m}$ do not approach the universal central extension.  This indicates that the
na\"ive lattice energy-momentum tensor is not the correct discretization of the continuum one.

The continuum field theory admits an improved (symmetric and traceless) energy-momentum tensor,
which differs from the canonical one by a total derivative.  Guided by this field-theory result, we
modify the lattice density by adding a lattice total-derivative (boundary) term,
\begin{align}
\tilde{h}_j
=h_j-it\left(c^\dagger_{j}c_{j}-c^\dagger_{j+1}c_{j+1}\right)
+(-1)^{j}i\delta\left(c^\dagger_{j}c_{j}+c^\dagger_{j+1}c_{j+1}\right).
\label{eq:hj_improved_SI}
\end{align}
With the periodic boundary condition, the added terms cancel in the sum, so that
$\sum_j \tilde h_j=\sum_j h_j$ and the energy spectrum is unchanged.  However, the higher Fourier
modes are modified, and it is precisely these modes that enter the Koo-Saleur construction.

\paragraph{Improved lattice Virasoro operators.}
We therefore construct the lattice Virasoro generators from the modified Hamiltonian density
$\tilde h_j$.  The corresponding momentum density and Fourier modes are defined by
Eqs.~\eqref{lattice_pj} and~\eqref{lattice_HnPn}, with $h_j$ replaced by $\tilde h_j$ throughout.
To simplify notation, in what follows we drop the tilde on lattice operators built from the modified
densities.

To evaluate the Virasoro generators explicitly, it is convenient to work in momentum space. Using the
Fourier transform~\eqref{eq:FT_theta} and the folded two-component spinor
$\Theta(p_m)=(\theta_{p_m-\pi/2},\,\theta_{p_m+\pi/2})^{\mathsf{T}}$, with
$p_m=2\pi m/N$ ($m=1,\dots,N/2$), the Koo-Saleur modes constructed from the improved density
$\tilde h_j$ take the quadratic form
\begin{align}
H_n
&=\frac{N}{2\pi v_F}\sum_{m=1}^{N/2}
\begin{pmatrix}
\theta^\dagger_{p_{m-n}-\pi/2} & \theta^\dagger_{p_{m-n}+\pi/2}
\end{pmatrix}
\mathcal{H}(p_m,p_n)
\begin{pmatrix}
\theta_{p_m-\pi/2} \\
\theta_{p_m+\pi/2}
\end{pmatrix},
\nonumber\\
P_n
&=\frac{iN}{2\pi v_F^2}\sum_{m=1}^{N/2}
\begin{pmatrix}
\theta^\dagger_{p_{m-n}-\pi/2} & \theta^\dagger_{p_{m-n}+\pi/2}
\end{pmatrix}
\mathcal{P}(p_m,p_n)
\begin{pmatrix}
\theta_{p_m-\pi/2} \\
\theta_{p_m+\pi/2}
\end{pmatrix},
\label{eq:HnPn_theta_SI}
\end{align}
where the single-particle matrices $\mathcal{H}$ and
$\mathcal{P}$ are
\begin{align}
\mathcal{H}(p_m,p_n)
&=4e^{i\frac{p_n}{2}}
\begin{pmatrix}
-t\cos\!\left(\frac{p_m-p_n}{2}\right)\sin\!\left(\frac{p_m}{2}\right)
&
-i\delta\cos\!\left(\frac{p_m}{2}\right)\cos\!\left(\frac{p_m-p_n}{2}\right)
\\[4pt]
-i\delta\sin\!\left(\frac{p_m}{2}\right)\sin\!\left(\frac{p_m-p_n}{2}\right)
&
t\sin\!\left(\frac{p_m-p_n}{2}\right)\cos\!\left(\frac{p_m}{2}\right)
\end{pmatrix},
\nonumber\\
\mathcal{P}(p_m,p_n)
&=-8i\left(t^2-\delta^2\right)e^{ip_n}\cos\!\left(p_m-\frac{p_n}{2}\right)
\begin{pmatrix}
\sin\!\left(\frac{p_m}{2}\right)\cos\!\left(\frac{p_m-p_n}{2}\right)&0\\[4pt]
0&\cos\!\left(\frac{p_m}{2}\right)\sin\!\left(\frac{p_m-p_n}{2}\right)
\end{pmatrix}.
\label{eq:HP_matrices_SI}
\end{align}

We next rewrite $H_n$ and $P_n$ in terms of the biorthogonal eigenmodes $\chi_{L/R}(p_m)$ and
$\eta_{L/R}(p_m)$ introduced in the previous subsection, together with the EP zero modes
$\chi_0$ and $\eta_0$.  For $n>0$, substituting Eq.~\eqref{eq:eigenmodes_def} into
Eq.~\eqref{eq:HnPn_theta_SI} yields the following expressions:
\begin{align}
H_n
&=\frac{N}{\pi}e^{i\frac{p_n}{2}}
\Bigg\{
\frac{\Delta}{\sqrt{2}}\sin\!\left(\frac{p_n}{2}\right)
\Big[\Delta_+\chi_R^\dagger(\pi-p_n)-\Delta_-\eta_R^\dagger(\pi-p_n)\Big]\eta_0
+\frac{\Delta}{\sqrt{2}}\sin\!\left(\frac{p_n}{2}\right)\chi_0^\dagger\Big[\chi_L(p_n)+\eta_L(p_n)\Big]
\nonumber\\
&\hspace{2.5em}
+\sum_{m=1}^{n-1}\sin\!\left(\frac{p_m}{2}\right)\sin\!\left(\frac{p_{m-n}}{2}\right)
\Big[\Delta_+\eta_R^\dagger(\pi+p_{m-n})\chi_L(p_m)
+\Delta_-\chi_R^\dagger(\pi-p_{m-n})\eta_L(p_m)\Big]
\nonumber\\
&\hspace{2.5em}
+\sum_{m=n+1}^{N/2-1}\sin\!\left(\frac{p_m}{2}\right)\cos\!\left(\frac{p_{m-n}}{2}\right)
\Big[\chi_R^\dagger(p_{m-n})\chi_L(p_m)-\eta_R^\dagger(p_{m-n})\eta_L(p_m)\Big]
\Bigg\},
\end{align}
\begin{align}
H_{-n}
&=\frac{N}{\pi}e^{-i\frac{p_n}{2}}
\Bigg\{
\frac{1}{\sqrt2}\cos\!\left(\frac{p_n}{2}\right)\chi_0^\dagger\Big[\Delta_-\chi_L(\pi-p_n)+\Delta_+\eta_L(\pi-p_n)\Big]
+\frac{\Delta}{\sqrt2}\cos\!\left(\frac{p_n}{2}\right)\Big[\chi_R^\dagger(p_n)+\eta_R^\dagger(p_n)\Big]\eta_0
\nonumber\\
&\hspace{2.5em}
+\sum_{m=1}^{N/2-n-1}\sin\!\left(\frac{p_m}{2}\right)\cos\!\left(\frac{p_{m+n}}{2}\right)
\Big[\chi_R^\dagger(p_{m+n})\chi_L(p_m)-\eta_R^\dagger(p_{m+n})\eta_L(p_m)\Big]
\nonumber\\
&\hspace{2.5em}
+\sum_{m=N/2-n+1}^{N/2-1}\sin\!\left(\frac{p_m}{2}\right)\sin\!\left(\frac{p_{m+n}}{2}\right)
\Big[\Delta_+\chi_R^\dagger(p_{m+n}-\pi)\eta_L(p_m)+\Delta_-\eta_R^\dagger(p_{m+n}-\pi)\chi_L(p_m)\Big]
\Bigg\},
\end{align}
\begin{align}
P_n
&=\frac{N}{\pi}e^{ip_n}
\Bigg\{
-\frac{1}{\sqrt{2}}\Delta\cos\!\left(\frac{p_n}{2}\right)\sin\!\left(\frac{p_n}{2}\right)
\Big[\Delta_+\chi_R^\dagger(\pi-p_n)-\Delta_-\eta_R^\dagger(\pi-p_n)\Big]\eta_0
+\frac{1}{\sqrt{2}}\cos\!\left(\frac{p_n}{2}\right)\sin\!\left(\frac{p_n}{2}\right)\chi_0^\dagger\Big[\chi_L(p_n)+\eta_L(p_n)\Big]
\nonumber\\
&\hspace{2.5em}
-\sum_{m=1}^{n-1}\sin\!\left(\frac{p_m}{2}\right)\sin\!\left(\frac{p_{m-n}}{2}\right)\cos\!\left(p_m-\frac{p_n}{2}\right)
\Big[\Delta_+\eta_R^\dagger(\pi+p_{m-n})\chi_L(p_m)-\Delta_-\chi_R^\dagger(\pi+p_{m-n})\eta_L(p_m)\Big]
\nonumber\\
&\hspace{2.5em}
+\sum_{m=n+1}^{N/2-1}\sin\!\left(\frac{p_m}{2}\right)\cos\!\left(p_m-\frac{p_n}{2}\right)\cos\!\left(\frac{p_{m-n}}{2}\right)
\Big[\chi_R^\dagger(p_{m-n})\chi_L(p_m)+\eta_R^\dagger(p_{m-n})\eta_L(p_m)\Big]
\Bigg\},
\end{align}
\begin{align}
P_{-n}
&=\frac{N}{\pi}e^{-ip_n}
\Bigg\{
\frac{1}{\sqrt2}\cos^2\!\left(\frac{p_n}{2}\right)\chi_0^\dagger\Big[\Delta_+\eta_L(\pi-p_n)-\Delta_-\chi_L(\pi-p_n)\Big]
+\frac{\Delta}{\sqrt2}\cos^2\!\left(\frac{p_n}{2}\right)\Big[\chi_R^\dagger(p_n)-\eta_R^\dagger(p_n)\Big]\eta_0
\nonumber\\
&\hspace{2.5em}
+\sum_{m=1}^{N/2-n-1}\sin\!\left(\frac{p_m}{2}\right)\cos\!\left(\frac{p_{m+n}}{2}\right)\cos\!\left(p_m+\frac{p_n}{2}\right)
\Big[\chi_R^\dagger(p_{m+n})\chi_L(p_m)+\eta_R^\dagger(p_{m+n})\eta_L(p_m)\Big]
\nonumber\\
&\hspace{2.5em}
+\sum_{m=N/2-n+1}^{N/2-1}\sin\!\left(\frac{p_m}{2}\right)\sin\!\left(\frac{p_{m+n}}{2}\right)\cos\!\left(p_m+\frac{p_n}{2}\right)
\Big[\Delta_-\eta_R^\dagger(p_{m+n}-\pi)\chi_L(p_m)-\Delta_+\chi_R^\dagger(p_{m+n}-\pi)\eta_L(p_m)\Big]
\Bigg\}.
\end{align}
The parameters appearing above are
$\Delta=-\frac{i\delta}{v_F},\ \Delta_+=\frac{v_F+2t}{2i\delta}$, and $\Delta_-=\frac{v_F-2t}{2i\delta}$.
Evaluating commutators among $H_n$ and $P_n$, and hence among
$L_n=(H_n+P_n)/2$ and $\bar L_n=(H_{-n}-P_{-n})/2$, we find that, in the scaling limit, the algebra
converges to a Virasoro algebra (more precisely, two commuting chiral Virasoro algebras) with
\begin{align}
c=-2.
\end{align}

\medskip

\paragraph{Indecomposability parameters on the lattice.}
For a logarithmic pair $(|\phi\rangle,|\psi\rangle)$ at conformal weight $h$, the LCFT definition of
the indecomposability parameter $\beta$ is
\begin{align}
A_-|\xi\rangle=|\phi\rangle,\qquad
A_+|\psi\rangle=\beta|\xi\rangle,\qquad
L_0|\psi\rangle=h|\psi\rangle+|\phi\rangle,
\label{eq:beta_def_SI}
\end{align}
where $A_-$ is a singular-vector operator constructed by Virasoro combination and $A_+$ is the adjoint of $A_-$.  In a finite-size lattice computation one typically finds non-unit
prefactors,
\begin{align}
A_-|\xi\rangle=x_1|\phi\rangle,\qquad
A_+|\psi\rangle=x_2|\xi\rangle,\qquad
L_0|\psi\rangle=h|\psi\rangle+x_3|\phi\rangle,
\end{align}
which can be brought to the canonical LCFT normalization by rescaling
$|\xi'\rangle=x_1^{-1}|\xi\rangle$ and $|\psi'\rangle=x_3^{-1}|\psi\rangle$.  This yields
\begin{align}
A_-|\xi'\rangle=|\phi\rangle,\qquad
A_+|\psi'\rangle=\frac{x_1x_2}{x_3}|\xi'\rangle,\qquad
L_0|\psi'\rangle=h|\psi'\rangle+|\phi\rangle,
\end{align}
and hence
\begin{align}
\beta=\frac{x_1x_2}{x_3}.
\label{eq:beta_x123_SI}
\end{align}

\paragraph{Level 1.}
The EP produces two fermionic zero modes $\chi_0$ and $\eta_0$, leading to a fourfold-degenerate
ground-state sector.  Using the right ground state defined in Eq.~\eqref{eq:GS_biorth},
\begin{align}
|\rho_{0}^R\rangle:=\prod_{m=1}^{N/2-1}\eta_R^\dagger(p_m)\,|0\rangle,
\end{align}
we introduce the other three states in the ground-state sector:
\begin{align}
|\phi_0^R\rangle=\chi_0^\dagger|\rho_{0}^R\rangle,\qquad
|\psi_0^R\rangle=\eta_0^\dagger|\rho_{0}^R\rangle,\qquad
|\xi_{0}^R\rangle=\chi_0^\dagger\eta_0^\dagger|\rho_{0}^R\rangle,
\label{eq:level0_states_SI}
\end{align}
together with the corresponding left states (biorthogonal duals),
\begin{align}
\langle \rho_{0}^L|:=\langle 0|\prod_{m=1}^{N/2-1}\eta_L(p_m),\qquad
\langle \phi_0^L|:=\langle \rho_{0}^L|\eta_0,\qquad
\langle \psi_0^L|:=\langle \rho_{0}^L|\chi_0,\qquad
\langle \xi_{0}^L|:=\langle \rho_{0}^L|\eta_0\chi_0.
\label{eq:level0_left_SI}
\end{align}
Within the subspace $\mathrm{span}\{|\phi_0^R\rangle,|\psi_0^R\rangle\}$, $L_0$ acts as a Jordan block:
\begin{align}
L_0\Big|_{\{\phi_0,\psi_0\}}
=
\begin{pmatrix}
h_0&\dfrac{\alpha_0N}{4\pi v_F}\\[6pt]
0&h_0
\end{pmatrix},
\qquad
h_0=-\frac{N}{4\pi}\cot\!\left(\frac{\pi}{N}\right)+\frac{N^2}{4\pi^2}-\frac{1}{12}
=0+\frac{\pi^2}{180N^2}+\mathcal{O}(N^{-3}),
\label{eq:L0_level0_SI}
\end{align}
where $\alpha_0=-4i\delta$~\eqref{zeromode_prefactor}.  In particular, in the scaling limit
all level-0 states have conformal weight $h=0$, while $|\psi_0^R\rangle$ is a generalized eigenstate
associated with $|\phi_0^R\rangle$.

The first excited logarithmic pair is obtained by acting with $\chi_R^\dagger(p_1)$:
\begin{align}
|\phi_1^R\rangle:=\chi_R^\dagger(p_1)|\phi_0^R\rangle,\qquad
|\psi_1^R\rangle:=\chi_R^\dagger(p_1)|\psi_0^R\rangle,
\qquad p_1=\frac{2\pi}{N}.
\label{eq:level1_states_SI}
\end{align}
In the basis $\{|\phi_1^R\rangle,|\psi_1^R\rangle\}$, $L_0$ again takes a Jordan form:
\begin{align}
L_0=
\begin{pmatrix}
h_1&\dfrac{\alpha_0N}{4\pi v_F}\\[6pt]
0&h_1
\end{pmatrix},
\qquad
h_1=h_0+\frac{N}{4\pi}\sin(p_1)+\frac{N}{8\pi}\sin(2p_1)
=1-\frac{299\pi^2}{180N^2}+\mathcal{O}(N^{-3}).
\label{eq:L0_level1_SI}
\end{align}
At level~1 the relevant singular-vector operators are $A_{1\pm}=L_{\pm1}$.  Let $|\xi_{1}^R\rangle := |\xi_{0}^R\rangle$, one finds
\begin{align}
L_{-1}|\xi_{1}^R\rangle
&=-\frac{N}{2\sqrt2\pi}\Delta\,e^{-i\frac{p_1}{2}}\cos\!\left(\frac{p_1}{2}\right)
\left[1+e^{-i\frac{p_1}{2}}\cos\!\left(\frac{p_1}{2}\right)\right]|\phi_1^R\rangle,
\nonumber\\
L_1|\psi_1^R\rangle
&=\frac{N}{2\sqrt2\pi}e^{i\frac{p_1}{2}}\sin\!\left(\frac{p_1}{2}\right)
\left[1+e^{i\frac{p_1}{2}}\cos\!\left(\frac{p_1}{2}\right)\right]|\xi_{1}^R\rangle,
\nonumber\\
L_0|\psi_1^R\rangle
&=h_1|\psi_1^R\rangle+x_3\,|\phi_1^R\rangle,
\label{eq:level1_actions_SI}
\end{align}
with $x_3$ given by the off-diagonal entry in~\eqref{eq:L0_level1_SI}.  Applying
Eq.~\eqref{eq:beta_x123_SI} gives
\begin{align}
\beta_1
=-\frac{5+3\cos(p_1)}{8p_1}\sin(p_1)
=-1+\frac{17}{48}p_1^2+\mathcal{O}(p_1^4)
=-1+\frac{17}{48}\left(\frac{2\pi}{N}\right)^2+\mathcal{O}(N^{-4}).
\label{eq:beta1_SI}
\end{align}

\paragraph{Level 2.}
At level~2 we consider
\begin{align}
|\phi_2^R\rangle:=\chi_R^\dagger(p_2)\chi_R^\dagger(p_1)|\phi_0^R\rangle,\qquad
|\psi_2^R\rangle:=\chi_R^\dagger(p_2)\chi_R^\dagger(p_1)|\psi_0^R\rangle,
\qquad p_2=\frac{4\pi}{N}.
\label{eq:level2_states_SI}
\end{align}
The $L_0$ action in the basis $\{|\phi_2^R\rangle,|\psi_2^R\rangle\}$ reads
\begin{align}
L_0=
\begin{pmatrix}
h_2&\dfrac{\alpha_0N}{4\pi v_F}\\[6pt]
0&h_2
\end{pmatrix},
\qquad
h_2=h_1+\frac{N}{4\pi}\sin(p_2)+\frac{N}{8\pi}\sin(2p_2)
=3-\frac{2699\pi^2}{180N^2}+\mathcal{O}(N^{-3}).
\label{eq:L0_level2_SI}
\end{align}
At this level the singular-vector operators take the form
$A_{2\pm}=L_{\pm1}^2-2L_{\pm2}$.  Let $|\xi_2^R\rangle:=\chi_R^\dagger(p_1)|\xi_{0}^R\rangle$,
we obtain
\begin{align}
(L_{-1}^2-2L_{-2})|\xi_2^R\rangle
&=-\frac{Ne^{-ip_1}}{\sqrt2\pi}\cos(p_1)
\left\{
\left[1+e^{-ip_1}\cos(p_1)\right]
+\frac{N}{8\pi}\sin(p_1)\left[1+e^{-i\frac{p_1}{2}}\cos\!\left(\frac{3p_1}{2}\right)\right]
\left[1+e^{-i\frac{p_1}{2}}\cos\!\left(\frac{p_1}{2}\right)\right]
\right\}
|\phi_2^R\rangle,
\nonumber\\
(L_1^2-2L_2)|\psi_2^R\rangle
&=\frac{Ne^{ip_1}}{\sqrt2\pi}\sin(p_1)
\left\{
\left[1+e^{ip_1}\cos(p_1)\right]
+\frac{N}{8\pi}\sin(p_1)\left[1+e^{i\frac{p_1}{2}}\cos\!\left(\frac{3p_1}{2}\right)\right]
\left[1+e^{i\frac{p_1}{2}}\cos\!\left(\frac{p_1}{2}\right)\right]
\right\}
|\xi_2^R\rangle,
\nonumber\\
L_0|\psi_2^R\rangle
&=h_2|\psi_2^R\rangle+x_3\,|\phi_2^R\rangle,
\label{eq:level2_actions_SI}
\end{align}
where $x_3$ is the off-diagonal entry in~\eqref{eq:L0_level2_SI}.  Using
Eq.~\eqref{eq:beta_x123_SI}, we find
\begin{align}
\beta_2
=&-\frac{N}{2\pi}\sin(2p_1)
\left\{
\left[1+e^{ip_1}\cos(p_1)\right]
+\frac{N}{8\pi}\sin(p_1)\left[1+e^{i\frac{p_1}{2}}\cos\!\left(\frac{3p_1}{2}\right)\right]
\left[1+e^{i\frac{p_1}{2}}\cos\!\left(\frac{p_1}{2}\right)\right]
\right\}
\nonumber\\
&\times
\left\{
\left[1+e^{-ip_1}\cos(p_1)\right]
+\frac{N}{8\pi}\sin(p_1)\left[1+e^{-i\frac{p_1}{2}}\cos\!\left(\frac{3p_1}{2}\right)\right]
\left[1+e^{-i\frac{p_1}{2}}\cos\!\left(\frac{p_1}{2}\right)\right]
\right\}
\nonumber\\
=&-18+\frac{125}{4}\left(\frac{2\pi}{N}\right)^2+\mathcal{O}(N^{-4}).
\label{eq:beta2_SI}
\end{align}

Therefore, the lattice computation agrees with the field-theory results~\eqref{eq:betaM-closed-form}.

\SMrestoretoc
\endgroup

\end{document}